\documentclass[a4paper,12pt]{article}
\pdfoutput=1
\usepackage[T1]{fontenc}
\usepackage{comment}
\usepackage{jcappub}
\usepackage{subcaption}
\usepackage{hyperref}
\usepackage{graphicx}
\usepackage{xcolor}
\usepackage{amssymb,amsmath,bm}
\usepackage[utf8]{inputenc}
\usepackage[export]{adjustbox}
\hypersetup{colorlinks=true, citecolor=blue}
\usepackage{multirow}
\usepackage{subcaption}
\usepackage{amsfonts}
\usepackage{siunitx}
\usepackage{booktabs}
\usepackage{mathtools}
\usepackage{placeins}
\usepackage{anyfontsize}
\usepackage{float}

\allowdisplaybreaks

\def\be{\begin{equation}}
\def\ee{\end{equation}}
\def\ba{\begin{eqnarray}}
\def\ea{\end{eqnarray}}

\newcommand\LCDM{$\Lambda$CDM}

\newcommand{\CITE}[1]{[{\bf CITE}]}
\definecolor{darkviolet}{rgb}{0.58, 0.0, 0.83}

\usepackage[outercaption]{sidecap}
\usepackage{subfiles}

\usepackage[toc, page]{appendix}
\usepackage{tikz}
\usetikzlibrary{calc, intersections, through, arrows, backgrounds}
\usetikzlibrary{arrows.meta}

\hypersetup{
    colorlinks=true,
    citecolor=blue,
    linkcolor=blue,
    linktoc=page
}

\usepackage{graphicx}

\usepackage[normalem]{ulem}

\title{\fontsize{20}{32}\selectfont{Nonlinear Scales in Luminal Horndeski - I. Halo
mass function and power spectrum boost in models with Vainshtein screening. \vspace{-0.2in} \\}}

\author[a]{\fontsize{15}{25}\selectfont Dani de Boe,}
\author[a]{Mattia Pantiri,}
\author[a,b]{Gen Ye,}
\author[a]{Alessandra Silvestri}

\affiliation[a]{Institute Lorentz, Leiden University, PO Box 9506, Leiden 2300 RA, The Netherlands}
\affiliation[b]{D\'{e}partement de Physique Th\'{e}orique, Universit\'{e} de Gen\`{e}ve,
24 quai Ernest-Ansermet, CH-1211 Gen\`{e}ve 4, Switzerland}

\emailAdd{\textcolor{blue}{deboe@lorentz.leidenuniv.nl}}
\emailAdd{\textcolor{blue}{pantiri@lorentz.leidenuniv.nl}}
\emailAdd{\textcolor{blue}{Gen.Ye@unige.ch}}
\emailAdd{\textcolor{blue}{silvestri@lorentz.leidenuniv.nl}}

\abstract{We investigate nonlinear structure formation in Horndeski gravity with luminal gravitational wave speed ($c_T = 1$), assuming Vainshtein screening within the spherical collapse model. We compute the critical and virial overdensities and evaluate the halo mass function. Building on the reaction approach, we present a  framework for the computation of the reaction and the resulting nonlinear matter power spectrum using the EFTofDE formulation of Horndeski gravity. We show results for the case of EFT functions that trace the evolution of  dark energy, and  specialize to the cubic galileon and nDGP models  for benchmarking against existing results. The framework interfaces with \texttt{EFTCAMB} for the linear evolution, though alternatives are possible. While restricted to Vainshtein-screened models, the current implementation focuses on qualitative trends and parameter dependencies. Further refinements and extensions to other screening mechanisms will be addressed in future work.}

\begin{document}

\maketitle

\section{Introduction}
\label{sec:introduction}

The large-scale structure (LSS) of the Universe — traced by the distribution of galaxies — offers key insights into the physics driving cosmic evolution~\cite{Peebles:1980lss}. As the Stage IV LSS surveys unfold, a central challenge in cosmology is to achieve a very accurate modeling of how initial perturbations in the early Universe grew into the complex structures we observe today, evolving from a linear regime to highly nonlinear, virialized systems. While a solid framework for this has been established in the Standard Model of Cosmology, 
\LCDM{}, across both regimes, a major goal of current and future cosmological surveys is to probe physics beyond \LCDM{}. Of particular interest to this work are the extensions of \LCDM{} associated to the nature of dark energy and of gravity at large scales.
Through the past decade, significant theoretical and numerical progress has been made in modeling the dynamics of background and linear perturbations in general extensions of \LCDM{},  including model-specific, parametric and non-parametric fits to available cosmological data. The nonlinear regime is less developed, but has attracted a significant effort in the past few years, as Stage IV LSS surveys are coming to life. In fact, these surveys crucially need to use the wealth of data that they will collect on smaller, nonlinear scales in order to reach the promised precision. Furthermore, small scales is where modified gravity effects, such as screening mechanisms, are expected to be most prominent. Over the past years, the community has steadily advanced toward this goal through the development of N-body codes for a range of modified gravity models (see~\cite{Winther:2015wfa} for an overview and comparison), faster hybrid methods such as \texttt{Hi-COLA}~\cite{Wright:2022krq}, and the powerful halo model reaction~\cite{Cataneo:2018cic,Bose:2020wch}. However, developing an efficient framework which enables to accurately explore the nonlinear matter power spectrum in  broad, parametrized beyond \LCDM{} scenarios remains  a key and timely challenge for guaranteeing robust tests of gravity on cosmological scales. 

In~\cite{Bose:2022vwi} the authors have made an important  effort in this direction. In this paper we make a similar step, with emphasis on the modeling of the halo mass function, treatment of (Vainshtein) screening and integration with \texttt{EFTCAMB}. We do so by extending the work of~\cite{Frusciante:2020zfs,Albuquerque:2024hwv} with the inclusion of the reaction approach and a running Planck mass. As such, our formalism has some key modeling differences compared to~\cite{Bose:2022vwi}. The integration with \texttt{EFTCAMB} will eventually enable the study of a broad class of covariant Galileons and Effective Field Theory of Dark Energy (EFTofDE) models. Currently, we have implemented only the Vainshtein screening, therefore that restricts the regime of applicability to models that display that mechanism. In future work, we plan to extend the framework to other classes of screening mechanisms and to scenarios involving multiple screening effects. We have also some work in preparation on incorporating Effective Field Theory of Large-Scale Structure (EFTofLSS) for the quasi-linear regime, in contrast to the SPT-based approach employed in \texttt{ReACT}. In another work in progress~\cite{deBoe}, we will validate the framework using N-body simulations. Thus, this paper is the first in a series devoted to the modeling of the nonlinear power spectrum in extended theories of gravity. The framework presented here constitutes an initial step toward an independent counterpart to the well-known \texttt{ReACT} code of~\cite{Cataneo:2018cic}, for which no alternative implementation currently exists.

We build a complete framework to calculate the nonlinear matter power spectrum for luminal Horndeski models, based on gravitational collapse and the halo model with a methodology heavily inspired by the \emph{reaction} approach introduced in~\cite{Cataneo:2018cic}. A key ingredient is the gravitational collapse of matter into virialized structures, the \emph{halos}. The spherical collapse model~\cite{Weinberg:2002rd} is one of the simplest yet most powerful analytical tools for studying this nonlinear mechanism, providing interesting physical insights. Initially developed within General Relativity (GR), it allows for a fully analytical treatment in an Einstein–de Sitter (EdS) Universe and a numerical one in the \LCDM{} scenario. More recently, the model has been extended to include screening mechanisms, i.e. the  characteristic mechanisms by which modified gravity theories reduce to GR in highly dense regions, enabling its application to a variety of modified gravity theories~\cite{Wang:1998gt,Horellou:2005qc,pace2010spherical,Pace_2012,Pace:2013pea,Pace:2018xqy,Pace:2019vrs,Pace:2017qxv,Frusciante:2020zfs,Albuquerque:2024hwv}. In this work, we focus on the broad class of Horndeski gravity theories~\cite{Horndeski:1974wa} with luminal gravitational waves, i.e $c_T=1$ as suggested by the direct detection of the multimessenger event GW170817~\cite{TheLIGOScientific:2017qsa,Baker:2017hug,LIGOScientific:2017zic,Goldstein:2017mmi}. Horndeski gravity  has gained significant attention in recent years as it offers a powerful framework for exploring the nature of gravity and dark energy on cosmological scales~\cite{Horndeski:2024sjk,Kobayashi:2019hrl}.

At the core of this work is a generalization of the spherical collapse model to luminal Horndeski gravity. We incorporate Vainshtein screening~\cite{Vainshtein:1972sx} to derive an effective gravitational constant — a key assumption we will examine in detail — and use the nonlinear matter density equations to track the evolution of a spherical overdensity through collapse. This allows us to estimate key quantities such as the \emph{critical density} contrast and the \emph{virial density}, which, as we will show, are model-dependent and generally differ from their \LCDM{} counterparts.

We then incorporate these results into a broader framework to compute the nonlinear matter power spectrum, using a methodology inspired by the \emph{reaction} approach of ~\cite{Cataneo:2018cic}. As a first step, we combine the outcomes of the spherical collapse model with the linear matter power spectrum predicted by the same theory to compute the halo mass function, which describes the abundance of halos as a function of mass~\cite{Press:1973iz,Sheth:1999su,Zhang:2005ar}. This serves as a fundamental ingredient of the halo model formalism~\cite{Seljak:2000gq,Cooray:2002dia}, which we use to derive the reaction function and, finally the nonlinear power spectrum. The spherical collapse model is a highly idealized description of structure formation. More realistic approaches account for the ellipsoidal nature of collapse through triaxial models. We incorporate this refinement at the level of the halo mass function by comparing the standard Press–Schechter formalism~\cite{Press:1973iz}, which assumes purely spherical collapse, with its generalizations: the ellipsoidal-collapse-based extension by Sheth and Tormen~\cite{Sheth:1999su} and the moving barrier formalism~\cite{Zhang:2005ar}. 

\textit{This work integrates the reaction formalism with the linear Einstein-Boltzmann solver \texttt{EFTCAMB}~\cite{Hu:2014oga,Raveri:2014cka}, enabling evaluation of the halo mass function and the nonlinear matter power spectrum for models of Horndeski gravity for which $c_T = 1$ and Vainshtein screening dominates.} The current implementation focuses on effective field theory parametrizations with $\alpha_i \propto \Omega_{\rm DE}(a)$, with the Cubic Galileon and nDGP included as additional benchmark models. Its extension to a broader class of luminal Horndeski theories and full validation will be the primary focus of a forthcoming study~\cite{deBoe}. We use this framework to study several aspects. We carry out a detailed comparison of the different models for the halo mass function in beyond \LCDM{} scenarios; for each case we investigate the differences in the halo mass function w.r.t. the \LCDM{} case, and how these propagate at the level of the reaction function and the nonlinear matter power spectrum. We compare the latter with differences that would be contributed by a simpler change in the cosmological parameters. 

The present work focuses on the highly nonlinear regime and a treatment of the halo mass function. Future developments will incorporate a perturbative treatment of the quasi-nonlinear regime through the ongoing implementation of one-loop corrections and the EFTofLSS in synergy with \texttt{EFTCAMB}~\cite{deBoe}. Furthermore, while Vainshtein is the most representative screening mechanism of Horndeski gravity, it is not the only possibility. In a work under preparation we are extending our formalism to more classes of screening~\cite{Pantiri}. 

The paper is structured as follows. In Section~\ref{sec:sp-el} we discuss the spherical collapse model and in Section~\ref{Sec:halo-mass} we introduce its relation to the halo mass function. Subsequently, the halo model \emph{reaction} is presented in Section~\ref{sec:NPK}. In Section~\ref{Sec:nonlinear} we study the nonlinear perturbations of Horndeski gravity with $c_T = 1$, under the  assumption of Vainshtein screening. The methodology is presented in Section~\ref{Sec:methods} and results will be discussed in Section~\ref{Sec:results}. In Section~\ref{Sec:conclusion} we draw the main conclusions and discuss future directions. 

\section{Spherical Collapse Model}
\label{sec:sp-el}

Let us begin by revisiting the spherical collapse model~\cite{Weinberg:2002rd}, without assuming GR as the underlying theory of gravity.
We focus on scalar perturbations to the Friedmann–Lemaître–Robertson–Walker (FLRW) line element, working in Newtonian gauge
\begin{equation}\label{eq:Newtonian_gauge}
    ds^2 = -(1+2\Psi)dt^2 + a^2(t)(1-2\Phi)\delta_{ij}dx^i dx^j\,,
\end{equation}
where $\Psi$ and $\Phi$ are the scalar perturbations to the metric and $a(t)$ is the scale factor. The starting point is to consider a matter density perturbation $\delta_m\equiv\delta\rho_m/\rho_m$ where $\rho_m$ is assumed to be the background energy density and the nonlinear continuity equation governing its time-dependence~\cite{Gunn:1972sv}:
\begin{equation}
    \ddot{\delta}_m + 2H \dot{\delta}_m - \frac{4}{3}\frac{\dot{\delta}_m^2}{1+\delta_m} = (1+\delta_m)\frac{\nabla^2 \Psi}{a^2} \,,     
\end{equation}
where the overdot denotes derivative w.r.t. time and $H=\dot{a}/a$. One then uses the Einstein equations to relate the Newtonian potential $\Psi$ to the matter perturbation. In the broad context of modified gravity, the Einstein equations will be modified, however, without loss of generality, they can be reduced to two non-dynamical, Poisson-like equations for the metric potentials, $\Phi$ and $\Psi$ ~\cite{Pogosian:2010tj}: 
\begin{align}
    \nabla^2 \Psi &= \frac{a^2}{2m_0^2}\mu\,  \rho_m \delta_m\,, \\
    \nabla^2 (\Psi+ \Phi) &= \frac{a^2}{m_0^2}\Sigma\, \rho_m\delta_m  \,,
\end{align}
where $m_0^2 = (8\pi G)^{-1}$ is the fixed Planck mass, with $G$ being Newton's gravitational constant, and any beyond \LCDM{} term is absorbed into two functions of time and scale, $\mu(a,\vec{x})$ and $\Sigma(a,\vec{x})$ which reduce to $\mu=\Sigma=1$ in the \LCDM \ limit. Notice that we focus on late times, when anisotropic stress of matter is negligible. These equations have been used extensively in the linear regime to explore constraints on gravity with cosmological surveys. On small but linear scales, working in Fourier space under the quasi-static approximation (where spatial gradients of metric potentials are neglected w.r.t. their time derivatives), one can derive simple analytical expressions for $\mu$ and $\Sigma$, which we will dub $\mu^{\mathrm{L}}$ and $\Sigma^{\mathrm{L}}$~\cite{Silvestri:2013ne}. However, incorporating information from smaller scales requires accounting for the nonlinear physics of screening mechanisms, which suppresses the fifth force associated with modifications to gravity in high-density environments (see~\cite{Brax:2021wcv} for a review on screening mechanisms).

In what follows, we denote the linear component of a given function with the superscript ${\mathrm{L}}$, while ${\mathrm{NL}}$ refers to the full, nonlinear expression and capture the nonlinear gravity relevant on smaller scales. While Horndeski gravity includes a broad range of scalar-tensor theories, these can be classified according to a handful of  screening mechanisms: Chameleon~\cite{Khoury:2003rn}, Symmetron~\cite{Hinterbichler:2010es}, Vainshtein~\cite{Vainshtein:1972sx}, and K-mouflage~\cite{Babichev:2009ee}. As detailed in~\cite{Lombriser:2016zfz}, the specific form of $\mu^{\mathrm{NL}}$ depends on the screening mechanism at the heart of the model. In Section~\ref{Sec:nonlinear}, we will outline the assumptions made in this work regarding the form of $\mu^{\mathrm{NL}}$.

We can now use the modified Poisson equation to get
\begin{equation}
    \ddot{\delta}_m + 2H \dot{\delta}_m - \frac{4}{3}\frac{\dot{\delta}_m^2}{1+\delta_m} = 4\pi G \mu^{\mathrm{NL}}\rho_m (1+\delta_m)\delta_m \,.
\end{equation}
The equation for the evolution of the linear matter density can easily be obtained from this expression by replacing $\mu^{\mathrm{NL}}$ with $\mu^{\mathrm{L}}$ and neglecting the quadratic term in $\delta_m$: 
\begin{equation}
\label{eq:delta_linear}
    \ddot{\delta}^L_m + 2H \dot{\delta}^L_m  = 4\pi G \mu^{\mathrm{L}}\rho_m \delta^L_m \,.
\end{equation}
We follow the collapse of a (spherically symmetric) mass $M$ defined by a top-hat profile with radius $R$; we assume that the total mass inside the radius is conserved during the collapse phase~\cite{Gunn:1972sv}, which implies
\begin{equation}
    M = \frac{4\pi}{3}\rho_m (1+\delta_m)R^3 = \mathrm{constant} \,.    
\end{equation}
Taking the second time derivative of this conservation equation, and combining it with the evolution equation for $\delta_m$ yields the following equation
\begin{equation}
    \frac{\ddot{R}}{R} = H^2 + \dot{H} - \frac{4\pi G}{3}\mu^{\mathrm{NL}}\rho_m \delta_m \,.  
\end{equation}
Introducing the new variable $y = R/R_i - a/a_i$, where $a_i$ is some initial scale factor and $R_i = R(a_i)$, we find
\begin{equation}
    y^{\prime \prime} = -\frac{H^\prime}{H}y^\prime + \left(1+\frac{H^\prime}{H}\right)y -\frac{\Omega_{m,0}H_0^2}{2a^3 H^2}\mu^{\mathrm{NL}}\delta_m \left(y+\frac{a}{a_i}\right) \,, 
\end{equation}
where the prime indicates the derivative w.r.t. $\mathrm{ln}(a)$ and $\Omega_{m,0}$ is the matter density parameter today. In case of a running Planck mass $M_\star$, we define the (non-relativistic) matter density parameter as $\Omega_m \equiv \frac{\rho_m}{3M_\star^2 H^2}$. Because of the mass conservation it must also hold that
\begin{equation}
    \delta_m = (1+\delta_{m,i})\left(1+\frac{a_i}{a}y\right)^{-3} - 1 \,,
\end{equation}
where $\delta_{m,i} = \delta_m(a_i)$. From the definition of $y$ it follows that $y_i = 0$ and in our analysis we will 
set $a_i \approx 6.66 \cdot 10^{-6}$ and $y_{i}^\prime = -\delta_{m,i}/3$ \footnote{From ref.~\cite{Frusciante:2020zfs}, it can be seen that this originates from $y_i^\prime = -\delta_{m,i}^\prime/(3(1+\delta_{m,i})) \approx -\delta_{m,i}^\prime/3$ and $\delta_m \propto \delta_m^\prime \propto a$ in the matter-dominated era.}. With these initial conditions the differential equation for $y$ is solved as an initial value problem using a shooting method to ensure collapse ($R=0$) at some provided scale factor (redshift) $a_c$ ($z_c$)~\cite{Bellini:2012qn}. In our numerical implementation we have the requirement $R/R_i < 1$ for collapse \footnote{It was verified that making the bound smaller has no significant impact on the results. It has also been shown that the procedure does not depend on the choice of $a_i$ as long as it is taken sufficiently small~\cite{Pace:2017qxv}.}. Once the collapse is solved, we can go back to the linear equation (\ref{eq:delta_linear}) and determine the value of $\delta^L_m$ at $z_c$, given the initial conditions identified above. Known as \emph{critical density}, denoted by $\delta_c$, this will be one of the main ingredients of the halo mass function. In simple terms, and as will be made clearer in Section~\ref{Sec:halo-mass}, it defines a threshold: when $\delta^L_m > \delta_c$, a halo is considered to have formed. 

Of course, in real astrophysical objects at some point the collapse will end before $R=0$ has been reached. This happens at the time of so-called virialization~\cite{Albuquerque:2024hwv}. The virialization occurs when the system reaches equilibrium so that the virial theorem is satisfied~\cite{Frusciante:2020zfs}:
\begin{equation}
\label{eq:vir}
T + \frac{1}{2}U = 0,    
\end{equation}
where $T$ and $U$ are the kinetic and potential energy given by
\begin{align}
    T = \frac{3}{10}M\dot{R}^2 \,, \,\,\,\,\,
    U = \frac{3}{5}(\dot{H}+H^2)MR^2 - \frac{3}{5}G \mu^{\mathrm{NL}}\frac{M}{R}\delta M \,.
\end{align}
The virialization scale factor $a_{\mathrm{vir}}$ is defined as the scale factor for which equation~(\ref{eq:vir}) is satisfied. The corresponding virial radius and mass are respectively defined as $R_{\mathrm{vir}} = R(a_{\mathrm{vir}})$ and  $M_{\mathrm{vir}} = \frac{4}{3}\pi \rho_{m,0}R_{\mathrm{vir}}^3 \Delta_{\mathrm{vir}}$, where the virial overdensity $\Delta_{\mathrm{vir}}$ is defined through~\cite{Frusciante:2020zfs}
\begin{equation}
\Delta_{\mathrm{vir}} = (1+\delta_m(a_{\mathrm{vir}}))\left(\frac{a_c}{a_{\mathrm{vir}}}\right)^3,    
\end{equation}
where $a_c$ is the scale factor at which the collapse occurs. 

In reality, the gravitational collapse of halos is not a spherically symmetric process and its description can be improved by considering an ellipsoidal configuration. A possible approach to include the ellipsoidal nature of collapse is through fast N-body simulations that solve the collapse equation in "ellipsoidal symmetry" directly (i.e. following the evolution of the three axes of the ellipsoid) \footnote{For this approach, see e.g.~\cite{Song:2021msd,Song:2023tnm}.}, which we will not rely on in our work. In this work we will rather stick to the approach in references~\cite{Frusciante:2020zfs,Albuquerque:2024hwv}, which is to solve the collapse in spherical symmetry and include the ellipsoidal nature of collapse via the \textit{halo mass function}. In the next section we will explain how this is obtained. The reason for this choice is that this approach is closer to the analytical approach to gravitational collapse. In Appendix~\ref{app:ellipsoidal} we provide a short summary of the concept of ellipsoidal collapse. 

\section{Halo Mass Function}
\label{Sec:halo-mass}

The gravitational collapse is closely related to the formation of structures in the Universe. In this section we will consider the halo mass function, which describes the number of halos within a comoving volume as a function of the mass of the halo and, as such, is a central concept in  observational cosmology. There are different formalisms to calculcate it, and here we briefly review them. The halo mass function affects gravitational lensing measurements~\cite{Weinberg:2002rd,Bartelmann:2010fz,Massey:2010hh}, constrains models of galaxy formation~\cite{White:1991gfh,Somerville:2014ika}, and informs predictions of halo merger rates~\cite{Lacey:1994su,Cohn:2000cm,Giocoli:2007gf,Ali-Haimoud:2017rtz,Fakhry:2020plg}. It also plays a key role in deriving cosmological constraints from clusters and large-scale surveys~\cite{Wang:1998gt,Majumdar:2002hd,Allen:2011zs,Planck:2013lkt,DES:2020mlx} and in studying the $\sigma_8$-tension~\cite{Gu:2023jef}. Given its broad relevance, understanding how the halo mass function is modified in scenarios beyond \(\Lambda\)CDM remains particularly informative.

The Press-Schechter (PS) function is a particular type of halo mass function that was introduced in~\cite{Press:1973iz} to predict the distribution of the number of halos as a function of their mass under the assumption of spherical symmetry. For this, one considers a scale $R$ over which the local density field $\delta_m$ is smoothed. The smoothed density field $\delta \equiv \delta_R$ is defined as:
\begin{equation}\label{eq:smoothed}
    \delta(a,\vec{x}) = \int d^3 \vec{y} \, W_R(|\vec{x} - \vec{y}|) \, \delta_m(a,\vec{y}) \,,    
\end{equation}
where $W_R$ denotes the window function. The variance of the smoothed density field is defined by
\begin{equation}\label{eq:variance}
    \sigma_M^2 = \frac{1}{2\pi^2} \int^\infty_0 dk \, k^2 \, W^2(kR)P_L(k,z) \,,    
\end{equation}
where the halo mass $M$ is defined via $M = \frac{4\pi}{3}\rho_{m,0}R^3$ and $W(kR)$ denotes the Fourier transform of the window function. We will assume a top-hat filter such that $W(x)=3x^{-3}(\sin(x)-x\cos(x))$. The Press-Schechter formalism provides the following function for the number of halos in some comoving volume~\cite{Press:1973iz}
\begin{equation}
\label{eq:ps}
\frac{dn}{dM}
=
\sqrt{\frac{2}{\pi}}\,
\frac{\rho_{m,0}}{M^2}\,
\frac{\delta_c}{\sigma_M}
\left|
\frac{d\ln \sigma_M}{d\ln M}
\right|
\exp\!\left(
-\frac{\delta_c^2}{2\sigma_M^2}
\right).
\end{equation}
where $\rho_{m,0}$ is the background matter density today, $\delta_c$ is the critical density, $M$ is the mass of the halo. 

The PS formalism clearly offers a simplified approach to the formation of halos, in which a region that exceeds the critical density threshold, known also as \emph{barrier}, is assumed to undergo spherical collapse and form a halo. The model relies on the Gaussian statistics of the initial density field and captures the general trend of hierarchical structure formation with low-mass halos being much more common than high-mass ones. Nowadays, the PS mass function can be understood as the spherical collapse limit of the \emph{excursion set formalism}~\cite{Bond:1990iw,Press:1973iz,Zhang:2005ar}. The latter is a refined model with a \emph{moving barrier} (MB), which accounts for the stochastic nature of the density field and incorporates ellipsoidal collapse dynamics, offering improved agreement with simulations. The key ingredients of the excursion set formalism are the variance (\ref{eq:variance}), which we will now denote with $S \equiv \sigma_M^2$, and the smoothed density field $\delta$. The formalism follows the growth of density fluctuations as the smoothing scale is varied; when we look at smaller regions, i.e. for smaller smoothing scales, $\delta$ fluctuates randomly due to the underlying Gaussian random field, resembling a random walk in the $\delta-S$ plane. The barrier, above which a region is expected to collapse, will be a curve in this plane and the \emph{first-crossing} of this barrier indicates at which (mass) scale collapse first occurs, and a halo is formed, for a given random walk. The distribution of first-crossing points gives the halo mass function—a statistical prediction for the abundance of gravitationally bound structures.

The case of spherical collapse can be viewed as a fixed barrier for which the probability distribution for first-crossing is Gaussian~\cite{Zhang:2005ar}. More generally, for the case of ellipsoidal collapse the barrier is instead a moving one $B(S)$~\cite{Sheth:2001dp}, i.e. it is not constant, and in that case the probability distribution for first-crossing is not Gaussian. The probability of first-crossing the barrier $B(S)$ between $S$ and $S+dS$ is defined as $f(S)dS$~\cite{Zhang:2005ar}, while $P(\delta,S)d\delta$ is defined as the probability that the random walk crosses between $\delta$ and $\delta + d\delta$ at $S$ without ever crossing the barrier before $S$~\cite{Zhang:2005ar}. Because of this definition, one has the following normalization condition
\begin{equation}
1 =
\int_0^S f(S')\,dS'
+ \int_{-\infty}^{B(S)} P(\delta,S)\,d\delta \,.
\end{equation}
In the absence of the barrier (i.e. $B(S) \rightarrow \infty$), it is assumed that $P(\delta,S) = P_0(\delta,S)$ is a Gaussian distribution defined by
\begin{equation}
    P_0(\delta,S) = \frac{1}{\sqrt{2\pi S}}\exp\left({-\frac{\delta^2}{2S}}\right) \,.    
\end{equation}
The halo mass function is directly related to the probability of first-crossing via the following equation~\cite{Zhang:2005ar}
\begin{equation}
\frac{dn}{dM} = \frac{\rho_{m,0}}{M}f(S)\left|\frac{dS}{dM}\right| \,.    
\end{equation}
Under the assumption that the barrier is linear ($B(S) = aS + b$ for some constants $a,b$) one recovers a Gaussian halo mass function~\cite{Press:1973iz,Zhang:2005ar}. The Press-Schechter halo mass function is understood as the case for which the barrier is constant and given by $B(S) = \delta_c$. For the ellipsoidal collapse instead one needs to solve  numerically the following set of equations to determine $f(S)$~\cite{Zhang:2005ar}
\begin{align}\label{eq:rw}
    f(S) &= g_1(S) + \int^S_0 dS^\prime \, f(S^\prime)g_2(S,S^\prime) \,, \\
    g_1(S) &= \left(\frac{B(S)}{S} - 2\frac{dB}{dS}\right)P_0(B(S),S) \,, \\
    g_2(S,S^\prime) &= \left(2\frac{dB}{dS} - \frac{B(S)-B(S^\prime)}{S-S^\prime}\right)P_0\left(B(S)-B(S^\prime),S-S^\prime\right) \,. 
\end{align}

The Sheth-Tormen description~\cite{Sheth:1999mn,Sheth:1999su,Zhang:2005ar} assumes that the barrier is of the form $B(S) = \sqrt{\tilde{a}}\delta_c [1+\beta(\tilde{a}\nu)^{-\alpha}]$, where $\nu \equiv \delta_c^2/S$, $\delta_c$ is the critical density from the spherical collapse model (which is a function of the redshift $z$), $\tilde{a} =  0.75$, $\beta = 0.485$ and $\alpha = 0.615$. These parameters were derived for $\Lambda$CDM, based on the approach in~\cite{Bond:1996ppp}. The halo mass function formulated in~\cite{Sheth:1999su}, as a fitting function based on Monte Carlo simulations, is of the form
\begin{equation}\label{eq:mps}
    \frac{dn}{dM} = \sqrt{\frac{2\tilde{a}}{\pi}}A \left[1 + \left(\frac{\tilde{a}\delta_c^2}{\sigma_M^2}\right)^{-p}\right]\frac{\rho_{m,0}}{M^2}\frac{\delta_c}{\sigma_M} \left|
\frac{d\ln \sigma_M}{d\ln M}
\right|\exp \left(-\frac{\tilde{a}\delta_c^2}{2\sigma_M^2}\right) \,,  
\end{equation}
where in this context $A = 0.3222$, $\tilde{a} = 0.75$ and $p = 0.3$ (following~\cite{Cataneo:2018cic} \footnote{Note that other references~\cite{Frusciante:2020zfs,Albuquerque:2024hwv} adopt the standard Sheth-Tormen values $A = 0.2162$, $\tilde{a} = 0.707$ and $p = 0.3$ instead.}). We refer to this as the modified Press-Schechter (MPS) halo mass function. 

The moving barrier formalism was included in the \texttt{CHAM} code, specialized to Hu-Sawicki $f(R)$ gravity~\cite{Hu:2017aei,Hu:2007nk}. Before concluding this part, let us  mention that in~\cite{Sheth:2001dp} a more general fitting formula for the first-crossing probability distribution function was proposed
\begin{equation}\label{eq:taylor}
    f(S) \, dS = \left|\sum^5_{n=0}\frac{(-S)^n}{n!}\frac{\partial^n B}{\partial S^n}\right|\exp \left(-\frac{B(S)^2}{2S}\right)\frac{dS}{S\sqrt{2\pi S}} \,.    
\end{equation}
Equation (\ref{eq:mps}) can thus be viewed as a truncated Taylor series in this respect. This ansatz for the first-crossing probability distribution has been compared to the moving barrier formalism in~\cite{Zhang:2005ar} showing good agreement at $z=0$ for $\nu>1$, while for $\nu<1$ the relative error grows to larger than ten percent. Equation (\ref{eq:taylor}) can thus be viewed as an approximation of equation (\ref{eq:rw}). In previous works~\cite{Frusciante:2020zfs,Albuquerque:2024hwv}  equation (\ref{eq:mps}) was used to compute the halo mass function under the assumption that the parameters $A,\tilde{a}$, $p$ for $\Lambda$CDM and modified gravity are the same. 
\begin{figure}[t]
    \centering    
    \includegraphics[width=0.5\linewidth]{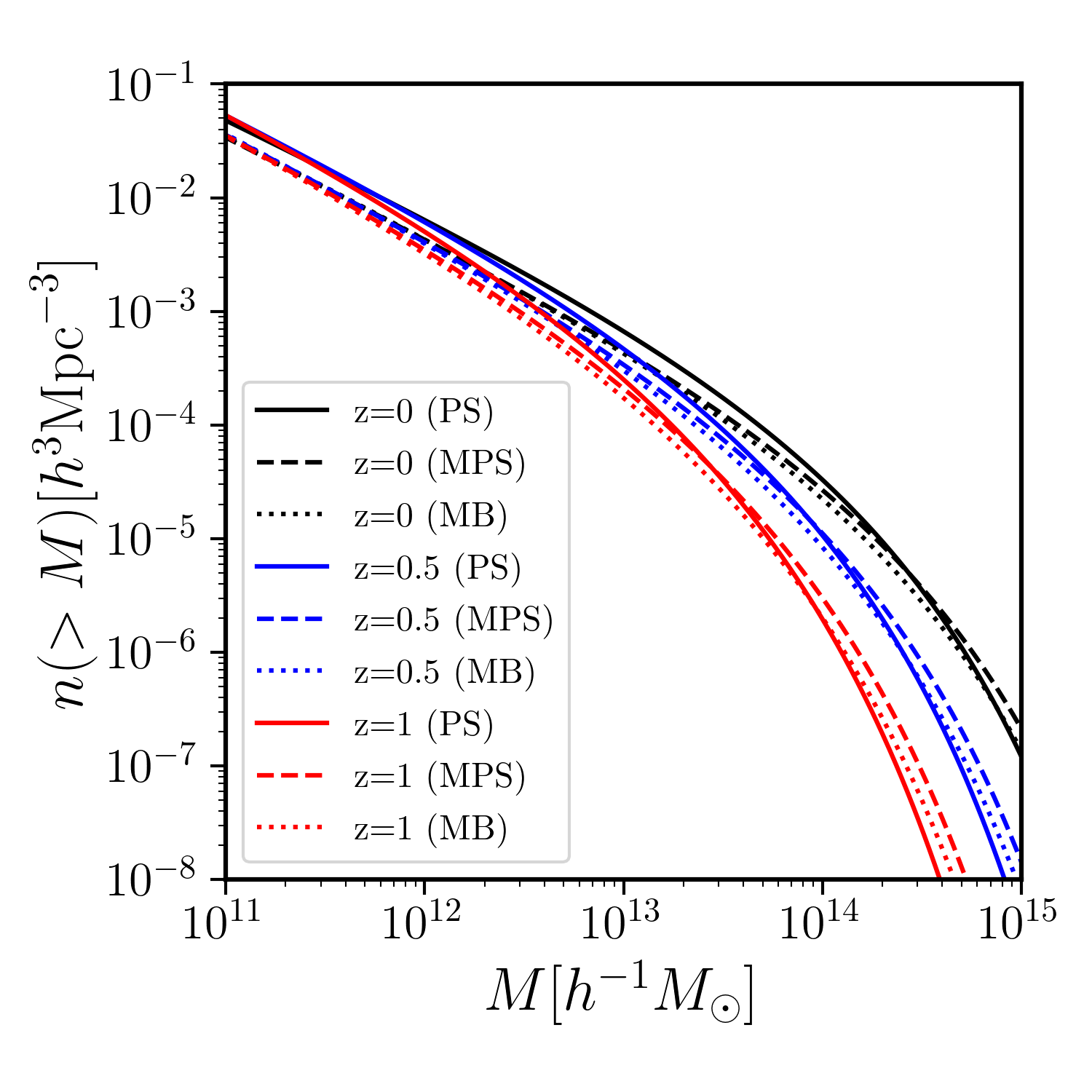}
    \caption{Predicted integrated halo mass functions $n(>M)$ for \LCDM{} at redshifts $z=0, \, 0.5, \, 1$ using the PS, MPS, and MB prescriptions.}
    \label{fig:lcdm_compare}
\end{figure}

Apart from the halo mass function, it is often useful to define the number density of halos above a given mass $M$ as \cite{Frusciante:2020zfs}:
\begin{equation}
n(>M) = \int^\infty_M \frac{dn}{dM^\prime} \, dM^\prime.    
\end{equation}
In Fig.~\ref{fig:lcdm_compare} we compare the number density of halos derived with the three different methods for the halo mass function, within the \LCDM{} framework. It can be noticed that, while PS and MB can differ significantly—PS overpredicts low-mass and underpredicts high-mass halos, MPS mainly underpredicts high-mass halos—MPS closely tracks MB for $M \lesssim 10^{14}\, h^{-1}\,M_\odot$, the range most relevant for calibration~\cite{Cataneo:2018cic}. These differences are found to be largely model-independent, justifying the use of \LCDM-calibrated MPS parameters $A$, $\tilde{a}$, and $p$ in modified gravity scenarios, with barrier and MG effects treated separately. 

\section{Nonlinear Matter Power Spectrum} 
\label{sec:NPK}

The halo mass function is a fundamental component of the halo model, a semi-analytic framework for describing the nonlinear clustering of matter~\cite{Cooray:2002dia}. The central assumption of this model is that all matter is bound within dark matter halos. Under this premise, the matter power spectrum can be decomposed into two contributions: the one-halo term, $P_{1h}(k)$, and the two-halo term, $P_{2h}(k) \sim P_{\rm L}(k)$. The one-halo term captures correlations of matter within individual halos and dominates on small scales, whereas the two-halo term accounts for correlations between distinct halos and governs large-scale behavior, asymptotically approaching the linear power spectrum.

Despite its versatility, the halo model exhibits some shortcomings, which have become increasingly relevant with the advent of Stage IV large-scale structure surveys. In particular, the model typically produces a non-smooth transition between the one-halo and two-halo regimes, often leading to an overestimation of power in the quasi-nonlinear range, and it is not inherently equipped to handle cosmologies beyond $\Lambda$CDM. To overcome these issues, a \emph{reaction} formalism has been developed in recent years~\cite{Cataneo:2018cic}. This approach introduces a reaction function that relates the nonlinear matter power spectrum of a modified cosmology to that of a reference model—usually $\Lambda$CDM—whose nonlinear spectrum is either known or more straightforward to compute with accuracy. In this work, we adopt this refined formalism to construct the nonlinear matter power spectrum for luminal Horndeski models.

We start from revisiting the halo model. Each halo is characterized by a halo mass function, a bias function (describing how halos trace the underlying matter distribution) and a profile function for the mass density. For the latter we adopt the common Navarro-Frenk-White (NFW) halo profile~\cite{1996ApJ...462..563N}:
\begin{equation}
\rho_h(r) = \frac{\rho_s}{(r/r_s)(1+r/r_s)^2},    
\end{equation}
where $r_s$ is related to the virial concentration via $c_{\mathrm{vir}} = R_{\mathrm{vir}}/r_s$ and $\rho_s$ is related to the virial mass through~\cite{Cooray:2002dia, Cataneo:2018cic}
\begin{equation}
\rho_s = \frac{M_{\mathrm{vir}}}{4\pi r_s^3}\left[\mathrm{ln}(1+c_{\mathrm{vir}})-\frac{c_{\mathrm{vir}}}{1+c_{\mathrm{vir}}}\right]^{-1}.    
\end{equation}
In \LCDM{} we adopt the following form for the virial concentration~\cite{10.1046/j.1365-8711.2001.04068.x}:
\begin{equation}
c_{\mathrm{vir}} = \frac{c_0}{1+z}\left(\frac{M_{\mathrm{vir}}}{M_\star}\right)^{-\alpha},    
\end{equation}
where $c_0 = 9$, $\alpha = 0.13$ and $M_\star$ is defined through $\nu(M_\star) = 1$. For the modified gravity models instead, we assume that the virial concentration takes the following form \cite{2004A&A...416..853D,Cataneo:2018cic} \footnote{In our analysis we will use $z = 100$ for the limit $z \rightarrow \infty$.}:
\begin{equation}
c_{\mathrm{vir}} = \frac{c_0}{1+z}\left(\frac{M_{\mathrm{vir}}}{M_\star}\right)^{-\alpha} \frac{g_{\mathrm{MG}}(z \rightarrow \infty)}{g_{\Lambda}(z \rightarrow \infty)},   
\end{equation}
where $z$ is the redshift and $g_{\mathrm{MG}}$, $g_\Lambda$ are the linear growth factors normalized to $z=0$ for the modified gravity theory and \LCDM{} respectively. The linear bias can be defined through \cite{Hu:2017aei}:
\begin{equation}
b_L(M_{\rm vir}) = 1 - \frac{\partial}{\partial \delta_c} \ln\!\left( \frac{dn}{dM_{\rm vir}} \right).
\end{equation}
The nonlinear power spectrum is then given by \cite{Cooray:2002dia}:
\begin{equation}
P(k,z) = I^2(k,z) P_{L}(k,z) + P_{1h}(k,z),    
\end{equation}
where $I(k,z),P_{1h}(k,z)$ are defined as
\begin{align}
I(k,z) &=
\int \mathrm{d}\ln M_{\mathrm{vir}}\,
\frac{\mathrm{d}n}{\mathrm{d}\ln M_{\mathrm{vir}}}
\left(\frac{M_{\mathrm{vir}}}{\rho_{m,0}}\right)
u(k,M_{\mathrm{vir}},z)\,
b_L(M_{\mathrm{vir}}),
\\[4pt]
P_{1\mathrm{h}}(k,z) &=
\int \mathrm{d}\ln M_{\mathrm{vir}}\,
\frac{\mathrm{d}n}{\mathrm{d}\ln M_{\mathrm{vir}}}
\left(\frac{M_{\mathrm{vir}}}{\rho_{m,0}}\right)^2
\left|u(k,M_{\mathrm{vir}},z)\right|^2 .
\end{align}
In these expressions $u(k,M_{\mathrm{vir}},z)$ refers to the Fourier transform of the NFW profile truncated at $R_{\mathrm{vir}}$ (see~\cite{Cooray:2002dia}) and normalized as $u(k \rightarrow 0,M_{\mathrm{vir}},z) = 1$ \footnote{In the numerical implementation, the limit $k \rightarrow 0$ is taken as $k=0.01 \, h \, \mathrm{Mpc}^{-1}$ (following \cite{Cataneo:2018cic}).} and $P_L(k,z)$ is the linear power spectrum.

The reaction approach builds heavily on the halo model, but improves it including a smooth interpolation between the one-halo and two-halo terms, as well as encoding cosmology dependence without resorting to refitted halo parameters. This is achieved through a \emph{halo-model reaction function}, which relates the nonlinear power spectrum of a given beyond-$\Lambda$CDM cosmology to that of a reference cosmology, or \textit{pseudo cosmology}. The latter is defined as a \LCDM{} model with initial conditions chosen such that its linear power spectrum matches that of the modified gravity or dark energy model at the \emph{target redshift} $z_f$, i.e., $P^{\mathrm{pseudo}}_{L}(k,z_f) = P^{\mathrm{MG}}_{L}(k,z_f)$
\footnote{In our work this will be taken as $z_f \in \{0,1\}$. In the remainder we will just write $z$ for $z_f$.}, where $\mathrm{MG}$ refers to the modified gravity theory under consideration, i.e. the real cosmology. Note that in general, the MG and pseudo linear power spectrum differ for $z \neq z_f$ (see ref.~\cite{Cataneo:2018cic} for more details). The reaction function builds on the halo model, introducing standard perturbation theory (SPT) for a smoother transition and two free parameters to capture the MG phenomenology \footnote{This form of the reaction assumes that there are no massive neutrinos. For the inclusion of massive neutrinos, see \cite{Cataneo:2019fjp}.}:
\begin{equation}
\label{eq:reaction}
\frac{P_{\rm NL}^{\rm MG}}{P_{\rm NL}^{\mathrm{pseudo}}} =\mathcal{R}(k,z)\equiv \frac{[(1-\mathcal{E})e^{-k/k_\star(z)} + \mathcal{E}(z)]P_L^{\mathrm{pseudo}}(k,z) + P_{1h}^{\mathrm{MG}}(k,z)}{P_L^{\mathrm{pseudo}}(k,z) + P_{1h}^{\mathrm{pseudo}}(k,z)}\,.   
\end{equation}
The phenomenological parameters $\mathcal{E}(z)$ and $k_\star(z)$ are defined through \cite{Bose:2024qbw}
\begin{align}\label{eq:reaction-quant}
    \mathcal{E}(z) &= \lim_{k \rightarrow 0} \frac{P_{1h}^{\mathrm{MG}}(k,z)}{P_{1h}^{\mathrm{pseudo}}(k,z)} \,, \\
    k_\star(z) &= -\bar{k}\left[\mathrm{ln}\left(\frac{A(\bar{k},z)}{P_{L}^{\mathrm{pseudo}}(k,z)} - \mathcal{E}(z)\right) - \mathrm{ln}(1-\mathcal{E}(z))\right]^{-1} \,,
\end{align}
with $\bar{k} = 0.06 \ h\,\mathrm{Mpc}^{-1}$ and $A(k,z)$ being defined as
\begin{equation}\label{eq:reaction-quant2}
A(k,z) = \frac{P_{1-\mathrm{loop}}^{\mathrm{MG}}(k,z) + P_{1h}^{\mathrm{MG}}(k,z)}{P^{\mathrm{pseudo}}_{1-\mathrm{loop}}(k,z) + P^{\mathrm{pseudo}}_{1h}(k,z)}(P^{\mathrm{pseudo}}_{L}(k,z) + P^{\mathrm{pseudo}}_{1h}(k,z)) - P_{1h}^{\mathrm{MG}}(k,z),
\end{equation}
where $1-\mathrm{loop}$ refers to the one-loop power spectrum defined within SPT. Notice that we have set $I(k)=1$ after confirming that including it in the reaction typically only introduces sub-percent level differences on the reaction $\mathcal{R}$, as shown also in~\cite{Cataneo:2018cic}.

The philosophy of the reaction approach is to separate the accurate modeling of nonlinear gravitational clustering from the modeling of MG. Computing nonlinear corrections directly in beyond $\Lambda$CDM cosmologies can be computationally expensive, whereas several well-tested tools exist to obtain nonlinear predictions in $\Lambda$CDM, such as the halo-fit prescription~\cite{2015MNRAS.454.1958M,10.1093/mnras/stw681}, $N$-body simulations and emulators. The reaction formalism exploits this by expressing the nonlinear power spectrum of a modified cosmology as a correction to that of a pseudo $\Lambda$CDM, thereby isolating the impact of new physics from the nonlinear modeling itself.
We aim to generalize the reaction formalism to encompass the full class of luminal Horndeski theories endowed with Vainshtein screening, yielding broadly applicable and computationally efficient predictions for the reaction function.

\section{Nonlinear Perturbations in Modified Gravity}
\label{Sec:nonlinear}

In this work we consider the subclass of Horndeski gravity with luminal tensor modes ($c_T=1$), which is described by the following action in Jordan frame~\cite{Horndeski:1974wa}
\begin{equation}
S = \int d^4 x \, \sqrt{-g} \, [G_2(\phi,X) - G_3(\phi,X)\Box \phi + G_4(\phi)R] + S_m \,,   
\end{equation}
where $g$ and $R$  are, respectively, the determinant and Ricci scalar of the metric $g_{\mu \nu}$, $G_2$, $G_3$ and $G_4$ are free functions of the field $\phi$ and its kinetic term $X \equiv -\frac{1}{2}\nabla_\mu \phi \nabla^\mu \phi$ \footnote{Note that our convention for $X$ and $G_3$ differ from the convention in ref.~\cite{Albuquerque:2024hwv}.} and $S_m$ is the matter action. The field equations can be found in~\cite{Kobayashi:2011nu}; in our case, specializing to $c_T=1$ ($G_{4,X}=0$, $G_5=0$) gives the following Einstein and field equations:
\begin{eqnarray}
\label{eq:einstein_eq}
    & -\frac{1}{2}G_{2X}\nabla_\mu \phi \nabla_\nu \phi - \frac{1}{2}G_2 g_{\mu \nu} + \frac{1}{2}G_{3X}\Box \phi \nabla_\mu \phi \nabla_\nu \phi + \nabla_{(\mu}G_3 \nabla_{\nu)}\phi - \frac{1}{2}g_{\mu \nu}\nabla_\lambda G_3 \nabla^\lambda \phi + \nonumber \\ 
    & + G_{4}G_{\mu \nu} +g_{\mu \nu}(G_{4\phi}\Box \phi - 2XG_{4\phi \phi}) - G_{4\phi}\nabla_\mu \nabla_\nu \phi - G_{4\phi \phi}\nabla_\mu \phi \nabla_\nu \phi =\frac{1}{2} T_{\mu \nu}^{(m)} \,, \\
     \nonumber\\
    & G_{2\phi} + \nabla_\mu G_{3\phi}\nabla^\mu \phi + G_{4\phi}R = \nabla^\mu [(-G_{2X} + 2G_{3\phi})\nabla_\mu \phi + G_{3X}\Box \phi \nabla_\mu \phi + G_{3X}\nabla_\mu X] \,,\nonumber\\\label{eq:field_eq} 
\end{eqnarray}
where we adopted the notation $G_{iX} = \partial_X G_i$ and $G_{i\phi}=\partial_\phi G_i$, and $T_{\mu \nu}^{(m)}$ is the energy-stress tensor of matter fields. 

We focus on scalar perturbations in Newtonian gauge, as defined in (\ref{eq:Newtonian_gauge}), and introduce the perturbation of the scalar field  via $\phi(t,\vec{x}) = \bar{\phi}(t) + \delta \phi(t,\vec{x})$ \footnote{When it is clear from the context, we will also write $\bar{\phi}(t)$ as $\phi(t)$.}. 
We can then expand eqs.~(\ref{eq:einstein_eq}) and (\ref{eq:field_eq}) in terms of $\Psi,\Phi$ and $\delta \phi$. Following~\cite{Frusciante:2020zfs}, we will apply the quasi static approximation (QSA), i.e. neglect time derivatives of fields w.r.t. spatial derivatives  (see~\cite{Sawicki:2015zya} for a discussion of its validity in modified gravity models),  and only keep the highest order derivatives of fields. 

Under these approximations, the nonlinear traceless Einstein equation, 00-Einstein equation
and scalar field equation give, respectively \footnote{In the notation of~\cite{Wright:2022krq} but with $\Psi \leftrightarrow \Phi$. The results have been cross-checked and were found to be in agreement.}
\begin{align}
    \nabla^2(2G_4(\Phi - \Psi) - A_1 Q) &=0 \,, \\
    2G_4 \nabla^2 \Phi &= \frac{a^2}{2}\rho_m \delta_m - A_2\nabla^2 Q \,, \\
    A_0 \nabla^2 Q - A_1 \nabla^2 \Phi - A_2 \nabla^2 \Psi + \frac{B_0}{a^2 H^2} Q^{(2)} &= 0, \,,
\end{align}
where $Q \equiv H\delta \phi/\dot{\phi}$ (overdot means derivative w.r.t. $t$), $Q^{(2)} \equiv (\nabla^2 Q)^2 - (\partial_i \partial_j Q)^2$ and the coefficients are functions of $G_2, G_3, G_4$ and their derivatives. In the following we will adopt the so-called $\alpha$-basis introduced in~\cite{Bellini:2014fua} in which the coefficients acquire the following expressions \footnote{The convention for $\alpha_B$ used in this work is $\alpha_B \equiv -\frac{1}{2}\alpha_B^{\mathrm{BS}}$, where $\alpha_B^{\mathrm{BS}}$ is the convention from ref.~\cite{Bellini:2014fua}.}:
\begin{align}
    A_1 &= M_\star^2 \alpha_M \,\,\,\,\,\,\,\,\,\,\,\,\,\,\,\,\,\,\,\,\,\,\,
    A_2 = -M_\star^2 \alpha_B \,\,\,\,\,\,\,\,\,\,\,\,\,\,\,\,\,\,\,\,\,\,\,
    B_0 = \frac{1}{2}M_\star^2 (-2\alpha_B + \alpha_M) \, \\
    A_0 &= \frac{M_\star^2}{H^2}\dot{H}(1+\alpha_B) - M_\star^2 \alpha_M (1-\alpha_B) + \frac{M_\star^2}{H}\dot{\alpha}_B + M_\star^2 \alpha_B + \frac{3}{2}M_\star^2 \Omega_m \,
\end{align}
For Horndeski gravity with $c_T^2=1$, the mapping between the $\alpha$-basis and the $G_i$ functions is given by:
\begin{align}\label{eq:alpha}
    &M_\star^2 = 2G_4 \,\,\,\,\,\,\,
    H M_\star^2 \alpha_M = \frac{d}{dt}M_\star^2 \,\,\,\,\,\,\, HM_\star^2 \alpha_B = -\dot{\phi}(XG_{3X} - G_{4\phi})\,\,\,\,\,\,\,\alpha_T = \alpha_H = 0\nonumber \\
    &H^2 M_\star^2 \alpha_K = 2X(G_{2X} + 2XG_{2XX} - 2G_{3\phi} - 2XG_{3X\phi})+ 12\dot{\phi}XH (G_{3X} + XG_{3XX})\,.
\end{align}
As we show in Appendix~\ref{app:calc} these equations can be used to determine an algebraic expression for  the phenomenological function $\mu^{\mathrm{NL}}$:
\begin{equation}\label{eq:munl_maintext}
    \mu^{\mathrm{NL}} = \frac{m_0^2}{M_\star^2} + 2 \left(\mu^L - \frac{m_0^2}{M_\star^2}\right)\left(\frac{R}{R_V}\right)^3 \left(\sqrt{1+\frac{R_V^3}{R^3}} - 1 \right)\,,
\end{equation}
where
\begin{equation}
    \frac{R^3}{R_V^3} = \left(\frac{H}{H_0}\frac{m_0}{M_\star}\right)^2 \frac{a^3}{\delta_m \Omega_{m,0}} \frac{(\alpha_M -\alpha_B)^3}{2(\alpha_M - 2\alpha_B)(m_0^2/M_\star^2-\mu^L)^2} \,.
\end{equation}
Physically, this ratio must satisfy $R^3 / R_V^3 \geq 0$, which implies $(\alpha_M - \alpha_B)(\alpha_M - 2\alpha_B) \geq 0$ in overdense regions ($\delta_m \geq 0$). This condition is satisfied by all models considered in this work. As a check of the calculations, it follows that indeed $\mu^{\mathrm{NL}}\approx \mu^{\mathrm{L}}$ for $R \gg R_V$. For the running Planck mass, we consider the early-time normalization $M_\star(z_{\mathrm{ini}}) = m_0$ consistent with CMB and BBN observations \footnote{We additionally considered imposing a late-time normalization 
\(M_\star(z=0)=m_0\). However, for the parametrization adopted in this work, 
such a normalization is not well-defined because such a model returns to \LCDM{} at early times in \texttt{EFTCAMB}.}. With the appropriate translation of conventions, the result from~\cite{Albuquerque:2024hwv} can be seen as a special case of (\ref{eq:munl_maintext}) in which $G_2,G_3$ are only functions of $X$ and $M_\star = m_0$ holds at all times. 

The largest caveat in the above calculation is that it assumes that Vainshtein screening plays the dominant role in the screening whereas the non-trivial $G_4(\phi)$ together with a $\phi$-dependence of $G_2$ can give rise to Chameleon type screening~\cite{Tsujikawa:2009yf}, and a non-trivial kinetic term, which is allowed in a generic form of $G_2(X,\phi)$, could exhibit a kinetic or K-mouflage screening~\cite{Gratia:2016tgq}. \emph{In fact we assume that the source and the environment are such that the Vainshtein mechanism is the dominant one}~\cite{Gratia:2016tgq}. This is reflected by that in our derivation of $\mu^{\mathrm{NL}}$ (eq.~(\ref{eq:munl_maintext})), we schematically assumed the derivative hierarchy $|\delta \phi| \ll |\partial_i \delta \phi| \ll |\partial_i^2 \delta \phi|$ and kept only the highest derivative terms. This has also been done implicitly in previous literature~\cite{Albuquerque:2024hwv}, but such an approximation depends really on the source properties such as the mass and also the energy scales involved~\cite{Gratia:2016tgq}. Because of the Vainshtein assumption, the framework is currently applicable to any covariant (Horndeski) model which displays Vainshtein mechanism, as well as to $\alpha_i(a)$-parametrizations derived  or adapted to such models. More generic phenomenological choices of the linear EFT functions $\alpha_i(a)$ do not uniquely determine a screening mechanism or nonlinear completion. In the current implementation we describe such models with an assumed Vainshtein-type nonlinear completion, rather than a prediction uniquely determined by the chosen $\alpha_i(a)$ \footnote{An alternative procedure for modeling the nonlinear regime would be the nPPF framework~\cite{Lombriser:2016zfz,Bose:2022vwi}.}.


\section{Methodology}
\label{Sec:methods}

With all the necessary theoretical ingredients in place, we now turn to investigating the clustering phenomenology of luminal Horndeski models. To this end, we set up a sophisticated end-to-end algorithm which \emph{samples the theory space} and, for each physically viable choice, derives predictions for the integrated halo mass function, the reaction function and, finally, calculates the nonlinear matter power spectrum. Before describing the structure of the algorithm, and the code implementing it, let us discuss the choices we make at the level of the theory. As already mentioned, while focusing on Galileon models, we work with the so-called $\alpha$-basis for EFTofDE, setting \footnote{Compared to~\cite{Albuquerque:2024hwv}, we include a non-zero $\alpha_M$ and explore the parameter space in $\alpha_M$ and $\alpha_B$, rather than restricting to the KGB model~\cite{Albuquerque:2024hwv}.}:
\begin{equation}\label{eq:param}
\alpha_i(a) = \alpha_{i,0} \, \Omega_{\mathrm{DE}}(a) 
\end{equation}
where $i \in \{K,B,M\}$. While different choices are possible, in this work we set $w_{\mathrm{DE}} = -1$, effectively fixing the background to a \LCDM{} one. For the models considered in this work, we set $\alpha_{K,0}=0.01$ in order to ensure that all scales relevant for the halo model calculation ($k \gtrsim 10^{-4}\, h\,\mathrm{Mpc}^{-1}$ \footnote{This range corresponds to the Fourier scales associated with the halo masses entering the halo model calculation.}) lie within the sound horizon, i.e. $k/aH > (c_s)^{-1}$ for $a>0.01$. This corresponds to a necessary but not sufficient condition for the validity of the QSA~\cite{Sawicki:2015zya}, which is assumed in the halo-model framework. It is possible to confirm that the models in fact obey QSA by directly comparing their full dynamics on linear scales with that obtained with a quasi-static evolution. We have performed this check for all models considered in this work and confirmed that $\mu$ agrees with its QSA counterpart, used in the halo-model calculation, at the sub-percent level for $k > 0.01\,h\,\mathrm{Mpc}^{-1}$ \footnote{Although the QSA counterpart of $\Sigma$ does not directly enter the spherical-collapse equations, we performed the same consistency check and found comparable agreement.}. Since this agreement deteriorates on large scales, and could marginally impact the final result, we performed an additional test following the approach of Ref.~\cite{Bose:2022vwi}. Namely, we compared the reaction and nonlinear power spectra computed using either the \texttt{EFTCAMB} linear power spectrum, where the dynamics is evolved fully, or a rescaled \LCDM{} one obtained under QSA. For the models studied here, this comparison showed sub-percent agreement at $k > 0.01\,h\,\mathrm{Mpc}^{-1}$.   

Parametrization (\ref{eq:param}) has been implemented in a new version of \texttt{EFTCAMB}~\cite{Hu:2014oga,Raveri:2014cka}, the public patch to the Einstein-Boltzmann solver \texttt{CAMB}~\cite{Lewis:1999bs,camb}. For a given set of $\{\alpha_{i,0}\}$, one can compute all quantities required for the halo mass function, reaction, and nonlinear power spectrum: $H/H_0$, $H'/H$, $\alpha_B$, $\alpha_B'$, $\alpha_M$, $P_L^{\rm MG}(k,z)$, $g(z)$, and $P_{\rm NL}^{\rm pseudo}(k,z)$, where the prime denotes a derivative w.r.t. $\ln (a)$. Unless explicitly mentioned otherwise, we assume the fiducial values for the cosmology to be~\cite{Planck:2018vyg} \footnote{We assume the best-fit \LCDM{} values for Planck 2018 TT,TE,EE+lowE+lensing+BAO coming from the \texttt{CamSpec} likelihood (see Table A.1. in~\cite{Planck:2018vyg}). The exact choice does not significantly alter the trends in $\alpha_{B,0}$ and $\alpha_{M,0}$. Fixing the MG model to have the same fiducial cosmology as \LCDM{} allows us to study the effects of $\alpha_{B,0}$ and $\alpha_{M,0}$ and cosmological parameters separately.}: $h=0.677$, $\Omega_{c,0}h^2 = 0.12$, $\Omega_{b,0}h^2 = 0.022$, $A_s = 2.1\cdot 10^{-9}$, $n_s = 0.967$, $\tau = 0.055$ and $\Omega_{r,0} = 8.51 \cdot 10^{-5}$. It was checked that the results of this work are not sensitive to the value of the optical depth $\tau$. Finally, all models considered in this work are explicitly checked for ghost and gradient stability conditions in \texttt{EFTCAMB}.

\subsection{The Framework} \label{sec:structure_code}
In order to obtain the desired predictions for the clustering functions, we construct a multifaceted code which builds on the spherical collapse model, \texttt{CHAM}~\cite{Hu:2017aei} \footnote{Parts of our code were taken and modified from the publicly available repository \url{https://github.com/hubinitp/CHAM}. These parts were also optimized for purpose of numerical speed.} and \texttt{ReACT}~\cite{Cataneo:2018cic}. The code is written in such a way that it connects seamlessly to \texttt{EFTCAMB}. The background evolution and linear cosmology are computed using \texttt{EFTCAMB}. At the current stage, only the $\alpha_i \propto \Omega_{\mathrm{DE}}$ parametrization is fully implemented; however, the framework is flexible and can readily accommodate alternative models. The code is also designed to incorporate models not originally supported by \texttt{EFTCAMB} such as nDGP. 

Using the background and linear cosmology in combination with the \textit{spherical collapse module}, one can compute the critical density $\delta_c$ and virialization parameters (e.g., $\Delta_{\mathrm{vir}}$) for collapse at a specified redshift. At the current stage, the spherical collapse module relies on the assumption of Vainshtein screening in calculating $\mu^{\mathrm{NL}}$. With some modifications the code could also handle models exhibiting  other types of screening. We are currently working on extending it to Chameleon-screened models. 

From the linear power spectrum provided by \texttt{EFTCAMB}, the variance of the smoothed density field $\sigma_M^2$ can be computed. By combining this with the critical density obtained from the spherical collapse module, the halo mass function can be determined. For the halo mass distribution, a NFW profile is assumed, whose Fourier transform can be calculated using the virialization parameters from the spherical collapse calculation. Together with the halo mass function, this allows computation of the 1-halo contribution to the power spectrum.

The procedure is then repeated for the \textit{pseudo cosmology}, in which a \LCDM{} background is assumed, but with its linear power spectrum at some target redshift being identical to that of the modified gravity theory. The spherical collapse calculation is performed again, generally yielding different values for the critical density and virialization parameters. Combining these results with the variance of the smoothed density field provides the halo mass function for the pseudo cosmology, from which the 1-halo power spectrum is computed using the Fourier transform of the NFW profile.

\begin{figure}[htbp]
    \centering
\includegraphics[width=1\linewidth]{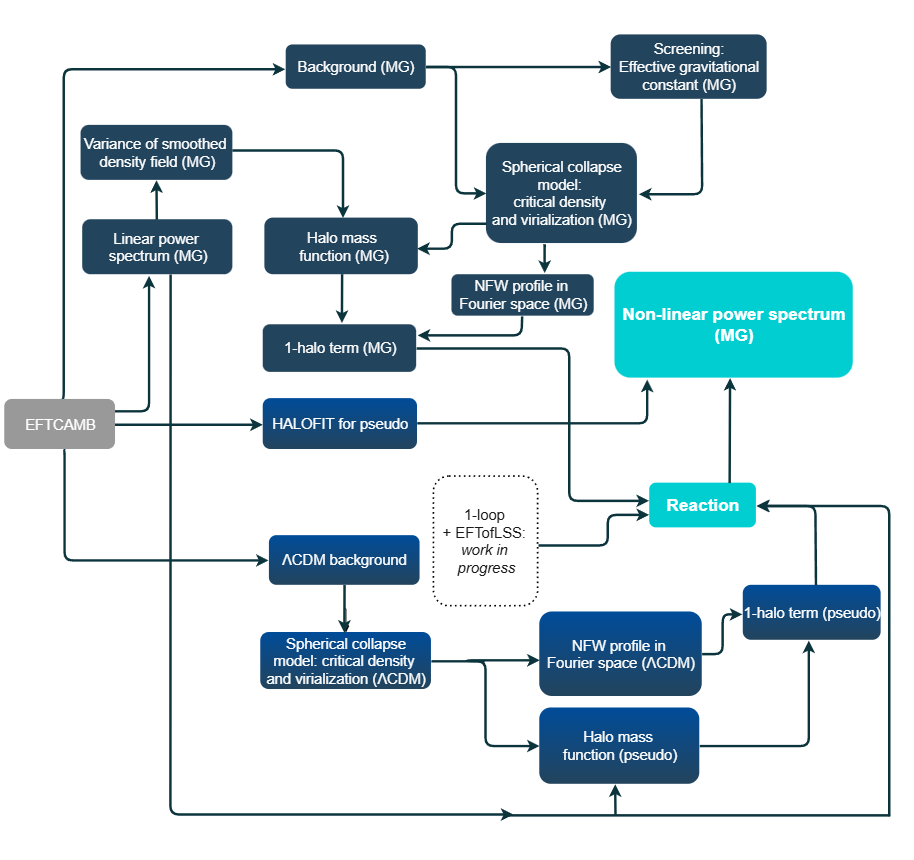}
    \caption{Flowchart of the algorithm used in this work. The label “MG” denotes the specific modified gravity theory considered.}
    \label{fig:flowchart}
\end{figure}

All the ingredients needed to compute the reaction are now in place. Figure~\ref{fig:flowchart} provides a detailed illustration of the code structure. At present, $\mathcal{R}$ is evaluated without one-loop contributions. We checked that ignoring the one-loop power spectra in eq.~(\ref{eq:reaction}) and then computing $k_\star$ generally yields a poor approximation; for instance, the reaction does not approach unity at small $k$. In the absence of one-loop terms, a better approximation is to put $k_\star \rightarrow \infty$. However, achieving sub-percent-level accuracy, especially on the quasi non-linear scales, would require a consistent, physically motivated determination of $k_\star$ (see also Figs.~2 and 3 in Ref.~\cite{Bose:2024qbw}), as we will address in forthcoming work using EFTofLSS~\cite{deBoe}. In Section~\ref{Sec:results} we will investigate the impact of $k_\star$, treating it as a free parameter over a physically motivated range, and compare this with the $k_\star \rightarrow \infty$ approximation. 

Having constructed the reaction, one can finally use it to obtain the nonlinear matter power spectrum of the modified gravity model from the nonlinear spectrum of the pseudo cosmology. Following the \texttt{ReACT} approach, we determine the   parameters for the latter by  enforcing a match between the pseudo and MG linear power spectra at the target redshift; we then compute the corresponding spectrum using \texttt{halofit}~\cite{2020A&A...641A.130M} within \texttt{EFTCAMB}. In future work~\cite{deBoe}, we plan to systematically compare \texttt{HMcode} and \texttt{halofit} for computing the nonlinear pseudo power spectrum, as their relative accuracy depends on the specific MG model~\cite{Atayde:2024tnr}.

\section{Results} 
\label{Sec:results}

Using the algorithm introduced in the previous section, we now explore the clustering phenomenology of Horndeski gravity models with luminal gravitational waves for some specific choices of the $\alpha$ functions. We use equation (\ref{eq:param}) and start with $\alpha_B$ only, eventually switching on also $\alpha_M$. The values of $\alpha_{B,0}$ and $\alpha_{M,0}$ are chosen for exploratory purposes to assess the \textit{qualitative} behavior of halo-model ingredients under parameter variations. No $N$-body simulations exist for these specific Horndeski models, so we focus on qualitative and relative trends rather than absolute $N$-body validation or direct observational inference. 

We present a detailed benchmarking against \texttt{ReACT} in App.~\ref{app:bench}. To ensure a consistent comparison, we adopt the same halo model ingredients as \texttt{ReACT}, including the MPS halo mass function, the concentration--mass relation, and the approximation $I(k)=1$. We also highlight differences between the two approaches and discuss directions for further refinement of our framework. Overall we find sub-percent agreement for the considered models, especially at the relevant scales ($k \lesssim 3\, h\, \mathrm{Mpc}^{-1}$). Of course, we use agreement with \texttt{ReACT}  as a numerical consistency check under these shared assumptions, not  as a physical validation of the nonlinear predictions, which would ultimately require dedicated $N$-body simulations~\cite{Cataneo:2018cic}.
Code-level agreement with \texttt{ReACT} is first established for common ingredients. Model-specific comparisons are then performed for Vainshtein-screened nDGP and cubic Galileon models. Finally, the implementation of $\alpha_i \propto \Omega_{\rm DE}$ is tested in the unscreened limit, showing consistency with the EFT time evolution and the reaction machinery, without testing the Vainshtein-type nonlinear extension.

When analyzing the results, it is important to consider the uncertainty related to the assumptions which we presented at length in the previous sections. Along with the halo model choices commonly adopted in the literature, and consistent with those used in \texttt{ReACT}, other sources of uncertainty in the current formulation of our framework are: the transition scale $k_\star$, the assumed Vainshtein-type nonlinear completion, the use of \texttt{halofit} for the pseudo nonlinear spectrum, and the lack of $N$-body calibration.

\subsection{Specific models with $\alpha_i \propto \Omega_{\rm DE}$}

\begin{figure}[h!]
  \centering
  \includegraphics[width=1\textwidth]{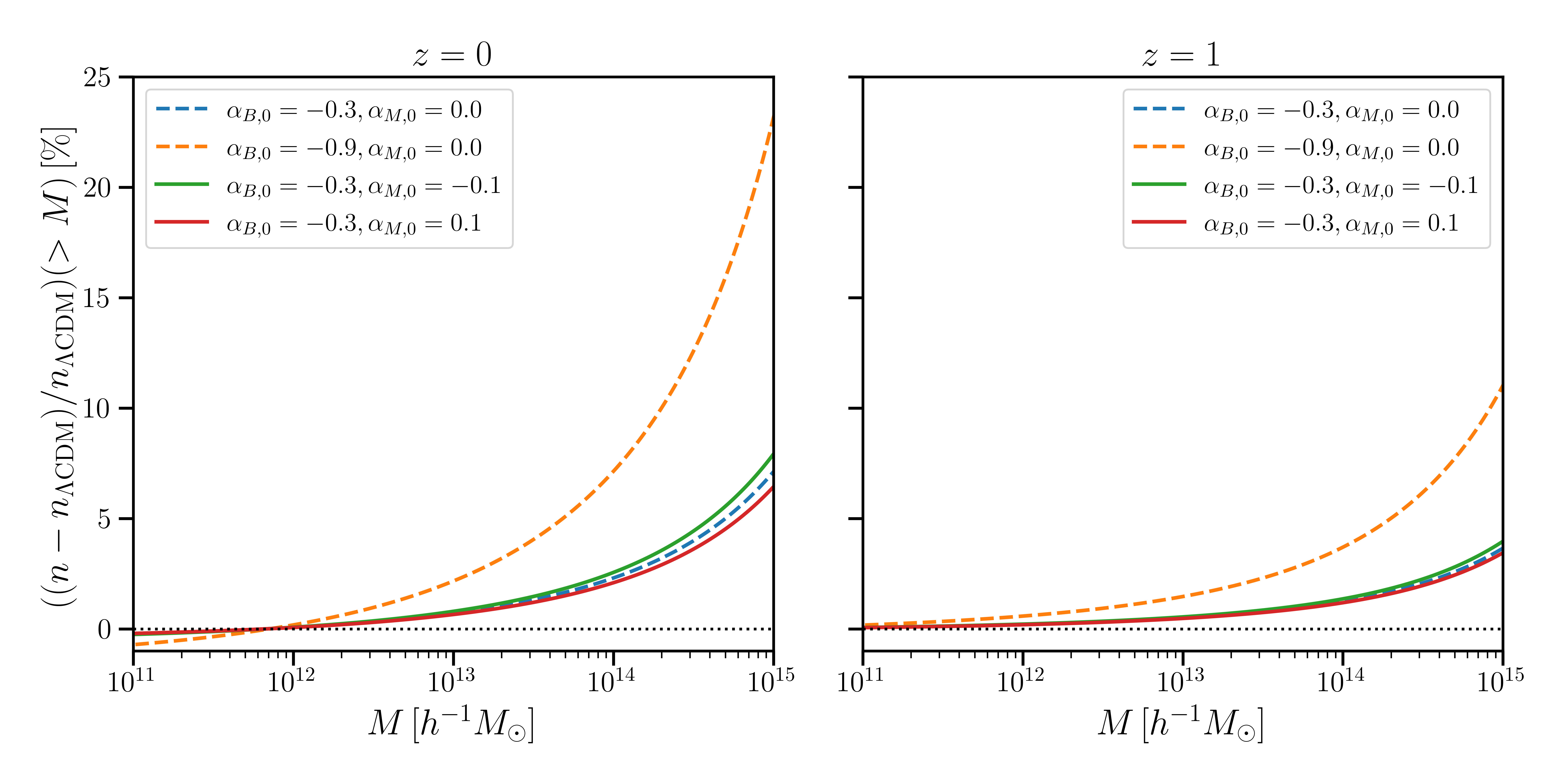}
\caption{Fractional difference in the integrated halo mass function $n(>M)$ for selected modified gravity models relative to \LCDM{} for two different redshifts. The halo mass functions are computed using the MPS formalism. The dotted line indicates $n = n_{\Lambda \mathrm{CDM}}$. Comparing the dashed lines  one can study the effect of a nonzero $\alpha_B$. Comparing the dashed blue line with the solid red and green lines, one can isolate the effects of a nonzero $\alpha_M$. Left panel: $z=0$; right panel: $z=1$.}
  \label{fig:HMF_specific}
\end{figure}

We begin the presentation of our results by comparing the integrated halo mass function of selected modified gravity models with the \LCDM{} one at two different redshifts ($z=0$ and $z=1$) in Figure~\ref{fig:HMF_specific}. We note that, for the parameter range considered, models characterized by a nonzero $\alpha_B$ generally increase the halo abundance relative to \LCDM{}. For $M \lesssim 10^{13} \, h^{-1} \, M_\odot$, the predictions remain close to \LCDM{}, while for $M \gtrsim 10^{13} \, h^{-1} \, M_\odot$, the enhancement becomes more pronounced. Larger values of $|\alpha_{B,0}|$ produce stronger deviations. The effect grows with time: at $z=0$, deviations reach $\sim 10$–$20 \%$, whereas at $z=1$ they are $\sim 1$–$10 \%$. Larger values of $|\alpha_{B,0}|$ produce stronger deviations. The effect grows with time, being more pronounced at low redshift than at high redshift. Including a negative (positive) $\alpha_{M,0}$ leads to an increase (decrease) in the halo abundance at large masses compared to $\alpha_{M,0} = 0$. We have verified that the choice of halo mass function formalism does not affect the qualitative trends reported in this work, particularly because we focus on the ratio relative to \LCDM{}, which remains qualitatively robust under different halo mass function prescriptions.
\begin{table}[htbp]
\caption{Spherical collapse parameters at redshift $z=0$ and $z=1$.}
\centering
\setlength{\tabcolsep}{4pt}
\begin{tabular}{|l|c|c|c|c|}
\hline
$(\alpha_{B,0},\alpha_{M,0})$ 
& $\delta_c(z=0)$ 
& $\delta_c(z=1)$ 
& $\Delta_{\rm vir}(z=0)$ 
& $\Delta_{\rm vir}(z=1)$ \\
\hline
$(0,0)$/\LCDM{}
& 1.676 & 1.684 & 330.0 & 199.9 \\
$(-0.3,0)$
& 1.684 & 1.686 & 318.7 & 198.5 \\
$(-0.9,0)$
& 1.706 & 1.689 & 291.9 & 195.5 \\
$(-0.3,-0.1)$ 
& 1.679 & 1.685 & 326.0 & 199.3 \\
$(-0.3,0.1)$ 
& 1.689 & 1.687 & 311.8 & 197.6 \\
\hline
\end{tabular}
\label{tab:scmg}
\end{table}

\begin{figure}[htbp]
    \centering
\includegraphics[width=0.6\linewidth]{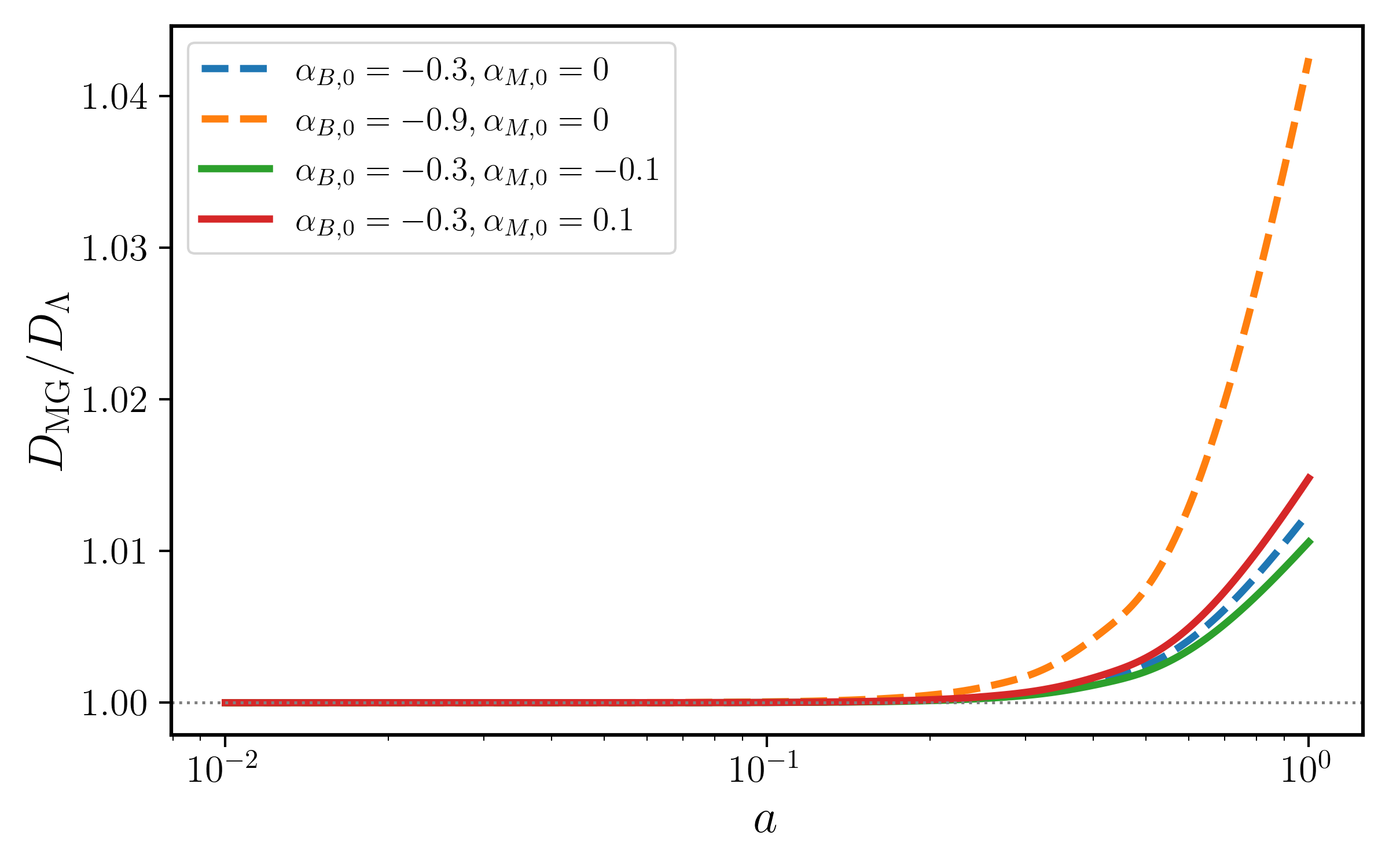}
    \caption{Relative linear growth factor $D_{\mathrm{MG}}/D_{\Lambda}$, with both growth factors normalized to $D(a=0.01)=0.01$. Comparing the dashed lines  one can study the effect of a nonzero $\alpha_B$. Comparing the dashed blue line with the solid red and green lines, one can isolate the effects of a nonzero $\alpha_M$.}
    \label{fig:linear_growth}
\end{figure}

Table~\ref{tab:scmg} lists the spherical collapse parameters $\delta_c$ and $\Delta_{\mathrm{vir}}$ at $z=0$ and $z=1$. Together with Figure~\ref{fig:linear_growth} for the linear growth, this helps interpret the trends seen in Figure~\ref{fig:HMF_specific}. For $\alpha_{M,0} = 0$, we see that the halo abundance at high masses increases with $|\alpha_{B,0}|$, due to the larger linear growth. On the other hand, the critical density $\delta_c$ is higher for $\alpha_{B,0}=-0.9$ than for $\alpha_{B,0} = -0.3$ at both $z=0$ and $z=1$. The net effect on the halo abundance, however, is dominated by the enhanced linear growth, which enters the variance of the smoothed density field $\sigma_M^2$, leading to an overall increase in halo numbers. When $\alpha_{M,0}$ is nonzero, the critical density at $z=0$ and $z=1$ increases for positive $\alpha_{M,0}$ and decreases for negative $\alpha_{M,0}$ relative to the $\alpha_{M,0}=0$ case. Although it is not displayed in Figure~\ref{fig:HMF_specific}, we checked that this conclusion holds for $\alpha_{B,0} = -0.9$, with the effect becoming slightly weaker with increasing $|\alpha_{B,0}|$. Figure~\ref{fig:linear_growth} shows that the linear growth is larger (smaller) for positive (negative) $\alpha_{M,0}$. Comparing with Figure~\ref{fig:HMF_specific}, we see that the (nonlinear) effect of $\delta_c$ dominates in these models: even though the linear growth changes as described, the halo mass function decreases (increases) for positive (negative) $\alpha_{M,0}$.

\begin{figure}[h!]
  \centering
  \includegraphics[width=1\textwidth]{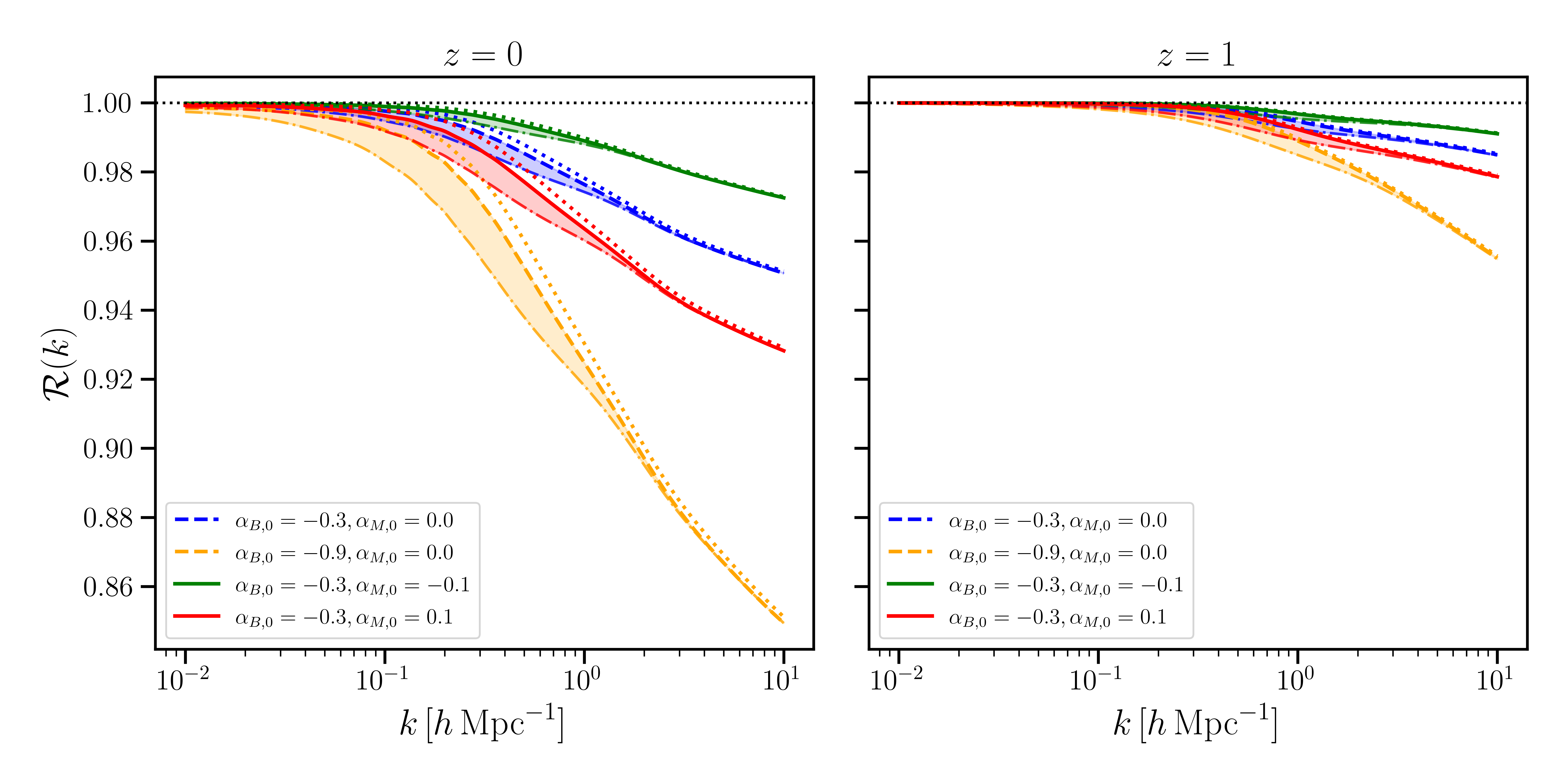}
\caption{Reaction $\mathcal{R}$ for selected modified gravity models relative to \LCDM{} for two different redshifts. The shaded regions indicate the $k_\star$ intervals, with dash-dotted lines corresponding to the lowest $k_\star$. Dotted lines show the $k_\star\rightarrow\infty$ approximation. Left panel: $z=0$; right panel: $z=1$.}
  \label{fig:Ri_vary_kstar}
\end{figure}

\begin{figure}[h!]
  \centering
  \includegraphics[width=1\textwidth]{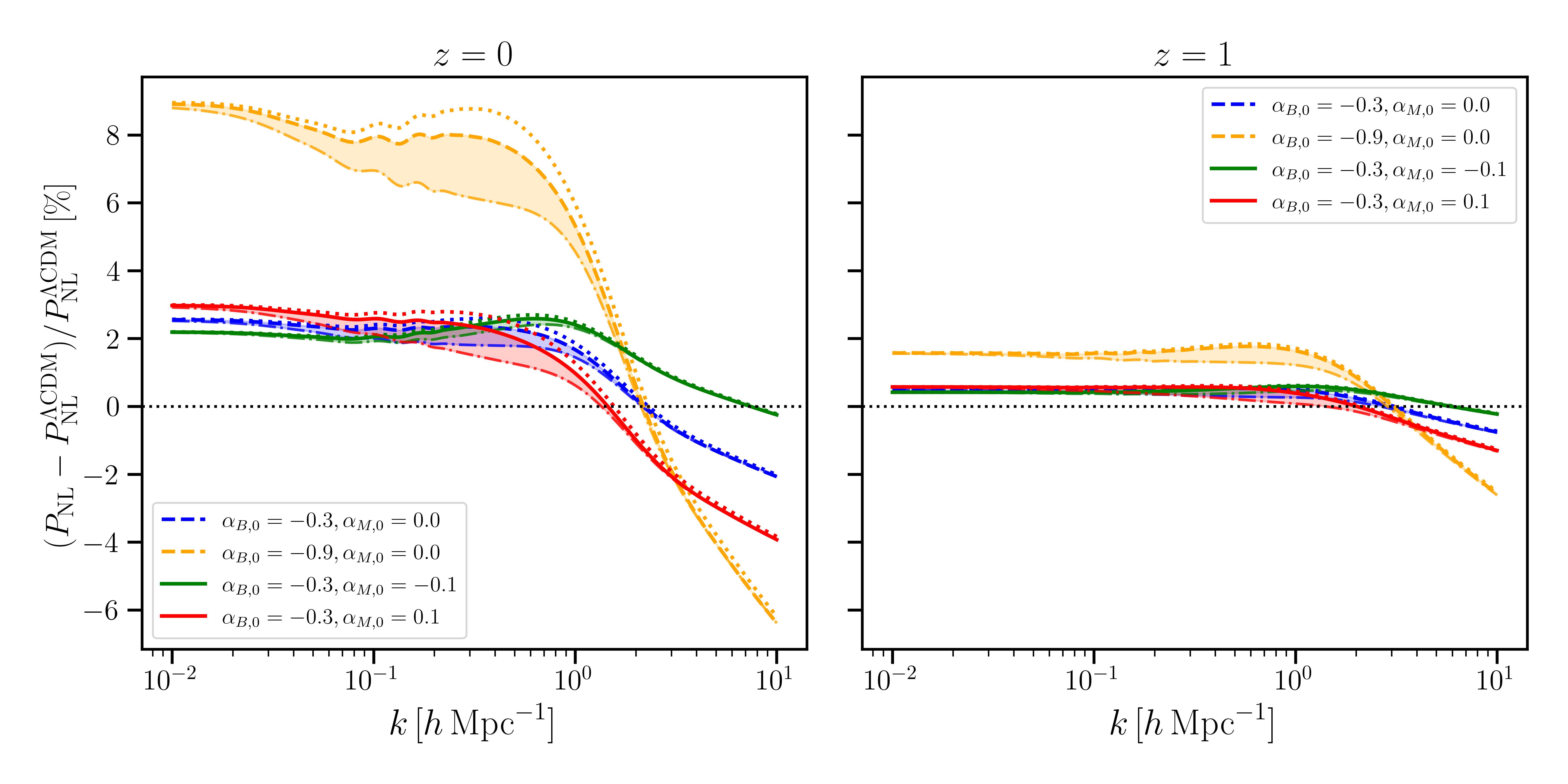}
    \caption{\emph{Boost factor}: Fractional difference of the nonlinear matter power spectrum $P_{\mathrm{NL}}$, computed via the reaction $\mathcal{R}$, with $P_{\mathrm{NL}}^{\Lambda \mathrm{CDM}}$. The shaded regions indicate the $k_\star$ intervals, with dash-dotted lines corresponding to the lowest $k_\star$. Dotted lines show the $k_\star\to\infty$ approximation. Left panel: $z=0$; right panel: $z=1$.}
  \label{fig:Pi_specific_kstar_pseudo}
\end{figure}

We now move on to the reaction and the nonlinear power spectrum. We show the results for the former in  Figure~\ref{fig:Ri_vary_kstar} and for the latter in~\ref{fig:Pi_specific_kstar_pseudo}, in the form of the boost factor $(P_{\rm NL}-P^{\rm LCDM}_{\rm NL})/P^{\rm LCDM}_{\rm NL}$. As mentioned before, in its current version our framework does not contain the SPT treatment of quasi nonlinear scales, which is necessary to fix $k_\star$. Because of this, we explore different options for $k_\star$: setting it to $k_\star\rightarrow\infty$, or varying it within physically motivated ranges $0.5 \lesssim k_\star(z=0) \lesssim 2 \, h\,\mathrm{Mpc}^{-1}$ and $1 \lesssim k_\star(z=1) \lesssim 10 \, h\,\mathrm{Mpc}^{-1}$. These $k_\star$ intervals cover the typical values in \texttt{ReACT} for Vainshtein-screened models such as nDGP and CG, and are comparable with previous studies~\cite{Cataneo:2018cic, Bose:2024qbw}, spanning from relatively strong nonlinear screening to effectively unscreened behavior on the scales relevant to this analysis.  As can be appreciated from Figures~\ref{fig:Ri_vary_kstar} and~\ref{fig:Pi_specific_kstar_pseudo}, for individual models, the approximation $k_\star \rightarrow \infty$ is sufficient to capture qualitative trends of the reaction and the nonlinear matter power spectrum. It also agrees with the finite-$k_\star$ results at the level of $\lesssim 2\%$ or better. It is important to keep in mind that while the ranges of $k_\star$ explored are well motivated, they are not guaranteed to encompass the true $k_\star$ value corresponding to the EFTofDE models under consideration. This thus remains a source of uncertainty which we will lift in the future versions of the framework.

In the current implementation we set $k_\star\to\infty$ by default, but the code can also accept a finite $k_\star$ as input. Moreover, should SPT results be available, they can be readily used to determine the model-specific value of $k_\star$. In general, $k_\star$ will have different values for different models, and this should be kept in consideration when carrying out a quantitative comparison of models on non-linear scales.

Looking at Figure~\ref{fig:Ri_vary_kstar}, it can be seen that the reaction decreases as $\alpha_{B,0}$ becomes more negative both for  $z=0$ and $z=1$. In contrast, the reaction increases (decreases) with more negative (positive) values of $\alpha_{M,0}$. These trends are explained by the combined effect of: linear growth factor (in $c_{\mathrm{vir}}$ and the halo mass function), critical density and virial overdensity. We found that the reaction $\mathcal{R}$ in the limit $k_\star \rightarrow \infty$ yields similar qualitative trends. As before, we can gain further insights via the information in Table~\ref{tab:scmg}. The latter shows that, both at $z=0$ and $z=1$, the critical density of the MG models are larger compared to \LCDM{}, while the virial overdensity is smaller. Moreover, we see from Table~\ref{tab:scmg} that the critical density, virial overdensity and linear growth of MG models are all closer to \LCDM{} at $z=1$ than at $z=0$. This explains why the reaction of MG models is closer to unity at all scales at $z=1$ compared to $z=0$. To understand the origin of the differences between the models, we fixed the spherical collapse parameters step by step across all models. We first imposed the same collapse threshold, $\delta_c$, and then also fixed $\Delta_{\mathrm{vir}}$, allowing us to isolate the effect of each quantity. We find that $\delta_c$ mainly drives the differences at intermediate scales, $0.1 \lesssim k \lesssim 1 \, h\,\mathrm{Mpc}^{-1}$, while $\Delta_{\mathrm{vir}}$ is responsible for most of the differences at smaller scales, $k \gtrsim 1 \, h\,\mathrm{Mpc}^{-1}$. When both $\delta_c$ and $\Delta_{\mathrm{vir}}$ are fixed, the differences between the $\alpha_{B,0}=-0.3$ models become very small. The remaining discrepancy for $\alpha_{B,0}=-0.9$ is mainly due to the different linear growth factor (see Fig.~\ref{fig:linear_growth}). Additionally setting the ratio of linear growth factors $g_{\mathrm{MG}}/g_{\Lambda}$ in the virial concentration to unity makes the reaction for the modified gravity models change by below sub-percent, showing that the effect of the ratio is sub-dominant compared to that of the linear growth factor (via the halo mass function) and the halo model parameters $\delta_c$ and $\Delta_{\mathrm{vir}}$ (see Table~\ref{tab:scmg}).

In conclusion, when $\alpha_{M,0}=0$, the enhancement of linear growth dominates, resulting in increased halo abundances and a reduced reaction overall. Compared to $\alpha_{M,0}=0$, positive (negative) $\alpha_{M,0} \neq 0$ yields a smaller (larger) $\Delta_{\mathrm{vir}}$ and a larger (smaller) $\delta_c$. The combined impact of these changes is to reduce both halo concentrations and halo formation efficiency, which lowers the reaction on small scales. For models with positive $\alpha_{M,0}$, the higher critical density together with the reduced virial overdensity suppresses halo formation (see Fig.~\ref{fig:HMF_specific}) and produces a smaller reaction relative to the $\alpha_{M,0}=0$ case in Figure~\ref{fig:Ri_vary_kstar}. Conversely, negative $\alpha_{M,0}$ lowers $\delta_c$ and increases $\Delta_{\mathrm{vir}}$, enhancing halo formation and concentration. The origin lies in that models with $\alpha_{M,0} \neq 0$ are associated with a running Planck mass (normalized at early time). A positive $\alpha_{M,0}$ decreases the effective gravitational strength at late times ($\mu^{\mathrm{NL}} \propto 1/M_\star^2$), which raises the critical density for collapse and lowers the virial overdensity. This slows structure growth and suppresses halo formation relative to the $\alpha_{M,0}=0$ case. Conversely, a negative $\alpha_{M,0}$ strengthens gravity, reducing $\delta_c$ and increasing $\Delta_{\mathrm{vir}}$, thereby enhancing late–time structure formation.

In Figure~\ref{fig:Pi_specific_kstar_pseudo}, we present the fractional difference of the nonlinear matter power spectrum, $P_{\mathrm{NL}}$, for several specific models relative to $P_{\mathrm{NL}}^{\Lambda \mathrm{CDM}}$ \footnote{Equivalently, Fig.~\ref{fig:Pi_specific_kstar_pseudo} can be interpreted in terms of the modified gravity boost $B(k,z) \equiv P_{\mathrm{NL}}/P_{\mathrm{NL}}^{\Lambda \mathrm{CDM}}$~\cite{Bose:2024qbw}.}. The influence of $\alpha_B$ and $\alpha_M$ on $P_{\mathrm{NL}}$ is scale-dependent. At both $z=0$ and $z=1$, $\alpha_B$ enhances $\Delta P_\mathrm{NL}$ for $k \lesssim 1 \, h\,\mathrm{Mpc}^{-1}$, with larger $|\alpha_B|$ producing stronger effects. For $k \gtrsim 1 \, h\,\mathrm{Mpc}^{-1}$, this trend reverses, as $\alpha_B$ lowers. Including a negative (positive) $\alpha_M$ leads to a decrease (increase) of $\Delta P_\mathrm{NL}$ at scales $k \lesssim 0.1 \, h\,\mathrm{Mpc}^{-1}$, while the trend reverses for $k \gtrsim 0.5 \, h\,\mathrm{Mpc}^{-1}$. Including a negative (positive) $\alpha_M$ generally decreases (increases) $\Delta P_\mathrm{NL}$ at large scales. In the intermediate regime, $0.1 \lesssim k \lesssim 0.5\, h\,\mathrm{Mpc}^{-1}$, the model dependence of the nonlinear matter power spectrum is sensitive to the choice of $k_\star$, and different models cannot be robustly distinguished without a physical determination of this parameter. At larger and smaller scales outside this interval, the difference between models becomes more stable against variations in $k_\star$. The behavior of $\Delta P_\mathrm{NL}$ in Figure~\ref{fig:Pi_specific_kstar_pseudo} is mostly explained by the the reaction of the different models in Figure~\ref{fig:Ri_vary_kstar}. Note that at large scales $\Delta P_\mathrm{NL} \neq 0$ due to the comparison of $P_\mathrm{NL}$ with a \LCDM{} power spectrum with the same cosmological parameters rather than with the  pseudo power spectrum. 

\begin{figure}[t]
    \centering
\includegraphics[width=0.5\linewidth]{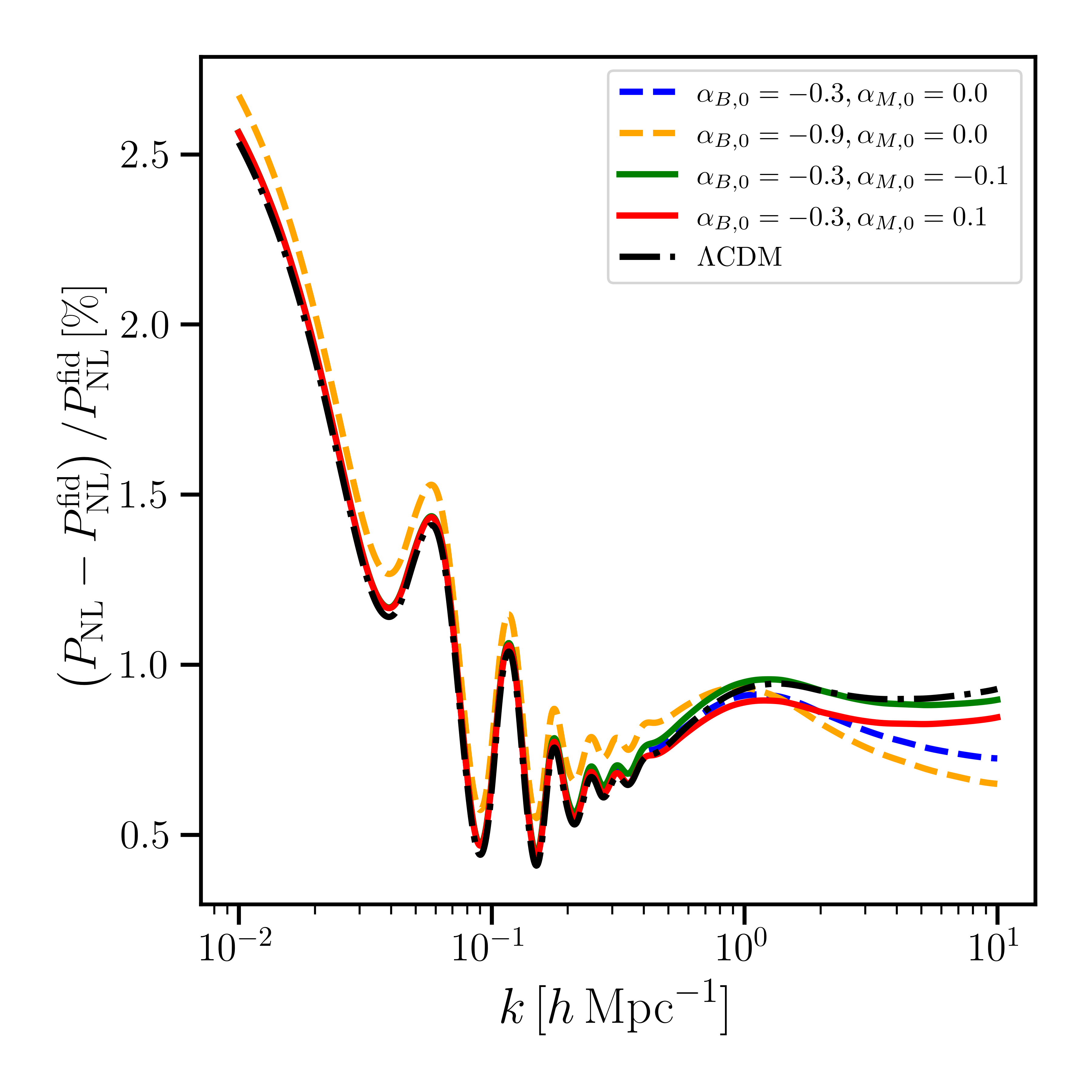}
\caption{Percent fractional difference between  nonlinear power spectra $P_{\mathrm{NL}}$ computed with the same cosmological parameters except for the value of $H_0$, for different models. For each model, the  fiducial spectrum $P_{\mathrm{NL}}^{\mathrm{fid}}$ is computed with the fiducial cosmology~\cite{Planck:2018vyg};  in $P_{\mathrm{NL}}$ instead the Hubble constant is increased by $1 \%$, while all other cosmological parameters remain fixed at their fiducial values. Dashed lines correspond to models with a nonzero $\alpha_B$ only, solid lines have a nonzero $\alpha_B$ and $\alpha_M$ and the dashed-dot line indicates \LCDM{}.}
\label{fig:H_specific}
\end{figure}
It is natural to ask how the effects of modified gravity compare with those induced by variations in cosmological parameters. If these effects are of comparable magnitude, accurate modeling of modified gravity becomes essential for precision cosmological inference. To this end, we investigate the sensitivity of the relevant functions to the values of the cosmological parameters at $z=0$. We compared the nonlinear matter power spectrum of the fiducial cosmology~\cite{Planck:2018vyg}, $P_{\mathrm{NL}}^{\mathrm{fid}}$, with that obtained by increasing the Hubble constant by $1 \%$, for \LCDM{} and our extended models. As can be seen in Figure~\ref{fig:H_specific}, for $k \lesssim 1 \, h\,\mathrm{Mpc}^{-1}$, deviations are largely model-independent, while at smaller scales ($k \gtrsim 1 \, h\,\mathrm{Mpc}^{-1}$) they reach $\sim 0.2 \%$, comparable to differences between models and halo mass function prescriptions. However, these high-$k$ scales are less relevant in practice, as the halo model is not expected to be reliable deep in the nonlinear regime. Additionally we have verified that varying $\Omega_{c,0}$, $\Omega_{b,0}$, $A_s$, or $n_s$ individually yields similar effects. Overall, modified gravity corrections are comparable in magnitude to those induced by small shifts in cosmological parameters, underscoring the importance of accounting for beyond-\LCDM{} effects in precision cosmological inference.

Finally, we have explored the impact of different halo mass function prescriptions (MB, PS, and MPS) on the reaction function. While these formalisms can differ by up to tens of percent—e.g., the PS approach overestimates the abundance of low-mass halos and underestimates that of high-mass halos (see Fig.~\ref{fig:lcdm_compare})—these discrepancies propagate relatively weakly to the reaction and nonlinear matter power spectrum. For modest parameter values, the resulting differences remain at the sub-percent level. However, for larger values of $|\alpha_{B,0}|$ or $|\alpha_{M,0}|$, which may be sampled in fits to data with broad priors, percent-level deviations can arise, particularly at low redshift. Overall, MPS provides a stable and validated choice, although full $N$-body calibration remains necessary to achieve sub-percent accuracy~\cite{Cataneo:2018cic}.

\section{Conclusions} \label{Sec:conclusion}
This work constitutes the first step in a broader program aimed at developing a reaction-based framework for predicting the nonlinear matter power spectrum in extended theories of gravity, with a particular focus on scalar--tensor models. 

The framework combines the spherical collapse model, the linear Einstein-Boltzmann solver for EFTofDE, \texttt{EFTCAMB}, and \texttt{CHAM} to enable the computation of the reaction $\mathcal{R}$ and the corresponding nonlinear power spectrum for EFTofDE models. In its current formulation, it is restricted to Horndeski models with luminal tensors and Vainshtein screening. We have benchmarked our implementation against \texttt{ReACT} for the CG and nDGP models, as well as for unscreened EFTofDE scenarios with braiding and a running Planck mass. For the specific models and tests considered in this work, this yields sub-percent agreement when using SPT one-loop results from \texttt{ReACT}, on the halo-model scales considered ($k \lesssim 3 \, h\,\mathrm{Mpc}^{-1}$). The $k_\star \rightarrow \infty$ approximation is mainly useful to capture qualitative trends of the reaction and the nonlinear matter power spectrum. For the specific tests performed, the approximation yields agreement at the level of $\lesssim 2\%$, but this should not be regarded as a general accuracy estimate. The framework is currently being extended in several directions, including the incorporation of additional screening mechanisms, numerical refinements, and one-loop EFTofLSS corrections. These developments are aimed at improving the robustness and precision of the predictions, ultimately enabling stringent comparisons with $N$-body simulations~\cite{deBoe}. More broadly, our goal is to provide a flexible and systematic tool for exploring nonlinear structure formation across a wide range of dark energy and modified-gravity scenarios. Such a capability will be increasingly important for efficiently probing large parameter spaces and fully exploiting the constraining power of forthcoming Stage IV cosmological surveys.

We have applied our framework to study the qualitative trends on nonlinear scales of stable quasi-static EFTofDE models parametrized by $\alpha_i = \alpha_{i,0} \, \Omega_{\mathrm{DE}}(a)$ ($i=B,M$). We find that a non-zero $\alpha_B$ increases halo abundances relative to \LCDM{}, with the effect being stronger for higher masses and more negative values of $\alpha_{B,0}$. On the other hand, a positive (negative) $\alpha_{M,0}$ results in a decrease (increase) in halo abundance. The reaction decreases with more negative $\alpha_{B,0}$, while it increases for negative $\alpha_{M,0}$ and decreases for positive $\alpha_{M,0}$. These trends arise from combined effects of linear growth and halo-model ingredients, and persist in the $k_\star \rightarrow \infty$ approximation. On the other hand, the nonlinear matter power spectrum deviates from \LCDM{} in a scale-dependent manner: increasing $|\alpha_{B,0}|$ generally enhances power at large scales while suppressing it at small scales. Positive (negative) values of $\alpha_{M,0}$ decrease (increase) power on large scales, whereas the trend reverses on small scales. Overall, these deviations are qualitatively smaller at early times than at late times. Finally, in the intermediate regime, $0.1 \lesssim k \lesssim 0.5 \, h \, \mathrm{Mpc}^{-1}$, the nonlinear matter power spectrum depends on the choice of $k_\star$ and cannot be interpreted in a fully model-independent way. Outside this regime, the predictions for $P_{\rm NL}$ are qualitatively stable against variations in $k_\star$. In contrast, the reaction shows a qualitatively stable behavior over a broad range of scales $k \gtrsim 0.01 \, h \, \mathrm{Mpc}^{-1}$ when comparing different models.
    
Looking ahead, several natural extensions arise. We are working on extending our framework to include Chameleon screening, so that the large, viable class of Generalized Brans-Dicke theories can be faithfully modeled. This is particularly important given the recent hints of non-minimally coupled gravity~\cite{Ye:2024ywg}. More intriguingly, Horndeski gravity and the EFTofDE contain a variety of operators capable of sourcing different screening mechanisms, depending also on the nature of the source being screened. A particularly promising direction is the investigation of double screening effects~\cite{Pantiri}, which may provide valuable insight into the interplay between multiple screening mechanisms in modified gravity theories. It will also be important to extend the reaction framework (eq.~(\ref{eq:reaction})) to include massive neutrinos~\cite{Cataneo:2019fjp} and baryonic physics~\cite{Bose:2021mkz}, which have a non-negligible impact on cosmology, particularly on small scales. Finally, determining the most appropriate form of the virial concentration $c_{\mathrm{vir}}$ in modified gravity remains an open question, and exploring alternative prescriptions, possibly informed by baryonic effects, could strengthen the robustness of our nonlinear predictions.

\acknowledgments
AS and DdB acknowledge support from the European Research Council under the H2020 ERC Consolidator Grant “Gravitational Physics from the Universe Large scales Evolution” (Grant No. 101126217 — GraviPULSE). AS and MP acknowledge support from the NWO and the Dutch Ministry of Education, Culture and Science (OCW), through ENW-XL Grant OCENW.XL21.XL21.025 DUSC. AS and GY acknowledge support from the NWO and the Dutch Ministry of Education, Culture and Science (OCW), through Grant VI.Vidi.192.069. We thank the developers of \url{https://app.diagrams.net/} for providing freely available software for creating flowcharts.
\appendix

\section{Ellipsoidal Collapse}
\label{app:ellipsoidal}

Chandrasekhar and Lebovitz~\cite{Chandrasekhar:1962psp} investigated the existence of stable homogeneous ellipsoids and obtained expressions for their gravitational potential. The gravitational collapse of homogeneous ellipsoids (also referred to as the ellipsoidal top-hat profile) has also been studied~\cite{Icke:1973fgc,White:1979gas,Peebles:1980lss,Barrow:1981gas}. In these approaches, the usual procedure is to generalize the spherical collapse model by replacing the single radius $R$ with three time-dependent axes $a_1,a_2,a_3$ \footnote{This notation should not be confused with the cosmological scale factor $a$.}, which evolve independently over time. The relationship between the axis ratios $a_1:a_2:a_3$ and the linear matter density $\delta_m^L$ was examined in~\cite{White:1993feg}. The analysis shows that, assuming the ellipsoidal top-hat profile remains uniform and ellipsoidal, the axis ratios tend to become increasingly extreme as the collapse progresses. 

The approaches in the references above differ from those in which an external tidal field is incorporated explicitly~\cite{Bond:1996ppp}. In this context, the time evolution of the principal axes of the strain tensor describes the ellipsoidal collapse under the Zeldovich approximation~\cite{Zeldovich:1970atl}. The ordering of the eigenvalues of the strain tensor determines the sequence of collapse along the principal axes. This description is necessary to capture the typical scenario in which an ellipsoidal overdensity first collapses along the shortest axis, then along the intermediate axis, and finally along the longest axis, ultimately forming a ‘pancake’ structure. Subsequently, filaments or clusters can form. This behavior is observed in numerical $N$-body simulations of large-scale structure and is therefore essential for accurately describing the gravitational collapse of halos. Correspondingly, the shear ellipticity and prolaticity of the ellipsoid are also time-dependent. In the formalism of~\cite{Bond:1996ppp}, the displacement field is decomposed into a background component and a fluctuation component. The background component is then smoothed over a characteristic large scale $R_{\mathrm{pk}}$ using a chosen filter (e.g., Gaussian or top-hat), with the precise implementation depending on the specific application of the theory. More recently, the formalism of~\cite{Bond:1996ppp} has been reformulated in a slightly different manner~\cite{Nadkarni-Ghosh:2014bwa}. Rather than describing the evolution of the axes $a_1,a_2,a_3$ as in~\cite{Bond:1996ppp}, this approach uses nine dimensionless parameters $\lambda_{a,i},\lambda_{d,i},\lambda_{v,i}$ (where $i \in \{1,2,3\}$)~\cite{Nadkarni-Ghosh:2014bwa}, which are functions of the axes and therefore provide an equivalent description of the ellipsoidal collapse.

The application of the ellipsoidal collapse formalism has recently been explored in the context of modified gravity theories using \texttt{PINOCCHIO}~\cite{Monaco:2001jg, Song:2021msd, Song:2023tnm}. Assuming Vainshtein screening as the dominant mechanism, ellipsoidal collapse can be studied within Lagrangian Perturbation Theory (LPT), enabling the computation of key astrophysical quantities such as the halo power spectrum and the halo mass function~\cite{Song:2021msd, Song:2023tnm}. This approach could be called the "fast $N$-body approach" as it relies on sampling different initial displacement fields and the corresponding matter density perturbations. Consequently, for a given initial matter density perturbation, there exist multiple equivalent initial conditions for the parameters $\lambda_{a,i},\lambda_{d,i},\lambda_{v,i}$. The results of this method are in good agreement with full $N$-body simulations while being significantly less computationally expensive~\cite{Song:2021msd, Song:2023tnm}.

\section{Calculations of Nonlinear Perturbations}
\label{app:calc}

From~\cite{Pogosian:2016pwr}, the large-scale limits of the phenomenological functions are given (under the stated approximations) by
\begin{align}\label{eq:mu_sig_QSA}
    \mu^L &= \frac{m_0^2}{M_\star^2}\left[1 + \frac{2}{c_s^2 \alpha}\left(\alpha_M -\alpha_B \right)^2 \right] \,, \\
    \Sigma^L &= \frac{m_0^2}{M_\star^2}\left[1 + \frac{1}{c_s^2 \alpha}\left(\alpha_M -\alpha_B \right)\left(\alpha_M -2\alpha_B \right)\right] \,, 
\end{align}
where $c_s^2$ is the scalar speed of sound and $\alpha = \alpha_K + 6\alpha_B^2$. To avoid ghost or gradient instabilities, both $c_s^2$ and $\alpha$ must be positive. Here $m_0^2 = (8\pi G)^{-1}$ is the fixed Planck mass, with $G$ the Newtonian gravitational constant. Moreover, the combination $\alpha c_s^2$ can be expressed as~\cite{Pogosian:2016pwr}
\begin{equation}
    \alpha c_s^2 = 2\left[\left(1+\alpha_B \right)\left(\alpha_M - \alpha_B - \frac{\dot{H}}{H^2}\right) - \frac{1}{H}\dot{\alpha}_B - \frac{3}{2}\Omega_m\right] \,.
\end{equation}
By combining the time-time (00) component of the Einstein equations with the traceless part, one obtains an equation for the potential $\Psi$. Eliminating $\Psi$ and $\Phi$ from the scalar field equation then leads to an expression involving only the scalar perturbation $Q$. Assuming spherical symmetry, such that $Q = Q(r)$ and $\delta_m = \delta_m(r)$, and following the procedure outlined in~\cite{Frusciante:2020zfs}, we arrive at
\begin{align}
    &\frac{1}{r^2} \frac{d}{dr}\left(r^2 \frac{dQ}{dr}\right) + \frac{2B_0}{a^2 H^2\Big(A_0 + \frac{1}{M_\star^2}A_2^2 + \frac{2}{M_\star^2}A_1 A_2\Big)}\frac{1}{r^2}\frac{d}{dr}\left(r\left(\frac{dQ}{dr}\right)^2\right) \nonumber \\
    &= \frac{A_1 + A_2}{M_\star^2 A_0 + A_2^2 + 2A_1 A_2}\frac{a^2}{2}\rho_m \delta_m \,.
\end{align}
Integrating the previous equation over $r$, we obtain
\begin{equation}
    r^2 \frac{dQ}{dr} + \frac{2B_0}{a^2 H^2\Big(A_0 + \frac{1}{M_\star^2}A_2^2 + \frac{2}{M_\star^2}A_1 A_2\Big)} r \left(\frac{dQ}{dr}\right)^2 = \frac{A_1 + A_2}{M_\star^2 A_0 + A_2^2 + 2A_1 A_2}a^2 m_0^2 G m(r) \,,
\end{equation}
where the enclosed mass is defined as
\begin{equation}
m(r) \equiv
4\pi \int_0^r \mathrm{d}r'\,
r'^2 \rho_m \delta_m \,.
\end{equation}
This is an algebraic equation for $dQ/dr$, which can be solved analytically. The solution reads
\begin{equation}
    \frac{dQ}{dr} = -\frac{a^2 H^2 \Big(A_0 + \frac{1}{M_\star^2}(A_2^2 + 2A_1 A_2)\Big)}{4B_0}r\left[1-\sqrt{1 + \frac{r_V^3}{r^3}} \, \right] \,,
\end{equation}
where the Vainshtein radius $r_V$ in this context is given by
\begin{equation}
    r_V \equiv \left[\frac{8B_0(A_1 + A_2)}{(A_0 + \frac{1}{M_\star^2}(A_2^2 + 2A_1 A_2))^2} \frac{Gm_0^2}{H^2 M_\star^2}m(r)\right]^{1/3} \,. 
\end{equation}
Specializing to a top-hat density profile, i.e. a constant $\delta_m$ within a sphere of radius $R$, one finds that $r_V \propto r$ and hence $dQ/dr \propto r$ (as shown in~\cite{Frusciante:2020zfs}). It is straightforward to verify that $Q^{(2)} = \frac{2}{3}(\nabla^2 Q)^2$. The scalar field equation then reduces to
\begin{equation}
    \nabla^2 Q + \frac{2B_0}{3a^2 H^2\Big(A_0 + \frac{1}{M_\star^2}(A_2^2 + 2A_1 A_2)\Big)} (\nabla^2 Q)^2 = \frac{A_1 + A_2}{A_0 + \frac{1}{M_\star^2}(A_2^2 + 2A_1 A_2)}\frac{a^2}{2M_\star^2}\rho_m \delta_m \,. 
\end{equation}
This algebraic equation again admits a Vainshtein-type solution. Evaluating it at the boundary of the overdensity, $r=R$, we obtain
\begin{equation}
    \nabla^2 Q\Big|_{r=R} = - \frac{3a^2 H^2 \Big(A_0 + \frac{1}{M_\star^2}\Big(A_2^2 + 2A_1 A_2\Big)\Big)}{4B_0}\left[1-\sqrt{1+\frac{R_V^3}{R^3}}\right] \,,
\end{equation}
where $R_V$ denotes the Vainshtein radius evaluated at $r=R$. In this case the enclosed mass satisfies $m(R) = \delta M$, and one finds that
\begin{equation}
    \frac{R^3}{R_V^3} = \left(\frac{H}{H_0}\frac{m_0}{M_\star}\right)^2 \frac{a^3}{\delta_m \Omega_{m,0}} \frac{(\alpha_M -\alpha_B)^3}{2(\alpha_M - 2\alpha_B)(m_0^2/M_\star^2-\mu^L)^2} \,.
\end{equation}
Rewriting $\nabla^2 Q|_{r=R}$ and substituting this expression into the equation for $\nabla^2 \Psi$, we find
\begin{equation}
    \nabla^2 \Psi = \frac{a^2}{2M_\star^2}\rho_m \delta_m \left[1 + 2 \frac{M_\star^2}{m_0^2}\left(\mu^L - \frac{m_0^2}{M_\star^2}\right)\left(\frac{R}{R_V}\right)^3 \left(\sqrt{1+\frac{R_V^3}{R^3}} - 1 \right)\right] \,.
\end{equation}
From this, the nonlinear effective gravitational coupling $\mu^{\mathrm{NL}}$ can be identified as
\begin{equation}\label{eq:munl}
    \mu^{\mathrm{NL}} = \frac{m_0^2}{M_\star^2} + 2 \left(\mu^L - \frac{m_0^2}{M_\star^2}\right)\left(\frac{R}{R_V}\right)^3 \left(\sqrt{1+\frac{R_V^3}{R^3}} - 1 \right)\,.
\end{equation}

\section{Benchmarking with \texttt{ReACT}}
\label{app:bench}

In the following, we benchmark our implementation against \texttt{ReACT} for the nDGP and Cubic Galileon (CG) models. In addition, we have implemented the $\alpha_i \propto \Omega_{\rm DE}$ model in \texttt{ReACT} and performed a comparison for the "unscreened" configuration in \texttt{ReACT}, defined by $\mu^{\mathrm{NL}} = \mu^{\mathrm{L}}$. Cosmological and model parameters are set to the values used in Ref.~\cite{Cataneo:2018cic}, ensuring a consistent comparison. We use the SPT result from \texttt{ReACT} at $\bar{k}=0.06\, h\,\mathrm{Mpc}^{-1}$ to determine $k_\star$ within our code and verify consistency with \texttt{ReACT}'s $k_\star$. The pseudo--cosmology parameters are stable across models and methods: at $z=0$ we find $\delta_{c,\rm pseudo} \approx 1.676$ and $\Delta_{\rm vir,pseudo} \approx 335$, while at $z=1$ we obtain $\delta_{c,\rm pseudo} \approx 1.684$ and $\Delta_{\rm vir,pseudo} \approx 200$, demonstrating sub-percent level agreement throughout with \texttt{ReACT}'s internal values and the ones reported in Ref.\cite{Cataneo:2018cic}. Overall, we find agreement between our implementation and \texttt{ReACT}, with differences in $\delta_c$ and $\Delta_{\rm vir}$ remaining at the sub-percent level and at most of order $\sim 1 \%$ across all models and redshifts considered.

In the following we review more in detail the comparison of each model considered. We start from the braneworld Dvali–Gabadadze–Porrati (DGP) model~\cite{Dvali:2000hr}, specializing to its normal branch, nDGP~\cite{Koyama:2005kd}. Following~\cite{Cataneo:2018cic}, we will assume a \LCDM{} expansion history for this model. The nDGP linear power spectrum is computed from its \LCDM{} counterpart by using the linear growth factors of nDGP and \LCDM{}. nDGP and its implementation in our code will be discussed in more detail in future work~\cite{deBoe}. For benchmarking we focus on the nDGPm ($r_c H_0 = 0.5$) and nDGPw ($r_c H_0 = 0.25$) models, where $r_c$ is the crossover scale (from 4D to 5D gravity) parameter~\cite{Koyama:2007ih,Schmidt_2010}. Table~\ref{tab:halo_parameters_full} and Figure~\ref{fig:nDGPm_nDGPw} show that we achieve sub-percent agreement on the halo model parameters and the reaction both with $k_\star \rightarrow \infty$ and by using \texttt{ReACT}'s SPT result to determine $k_\star$.

\begin{table*}[htbp]
\caption{Comparison of halo-model parameters for nDGPm and nDGPw at $z=0$ and $z=1$. Here $\delta_c$ is the linear collapse threshold, $\Delta_{\rm vir}$ is the virial overdensity, $k_\star$ is the characteristic transition scale, and 
$\mathcal{E} = P_{1h}^{\rm MG} / P_{1h}^{\rm pseudo}(k\rightarrow 0)$. The \texttt{ReACT} values of $\delta_c$ have been converted to our MG convention, and $k_\star$ in our case is computed from the SPT result at $\bar{k} = 0.06 \, h\,\mathrm{Mpc}^{-1}$.}
\centering
\begin{tabular}{|c c c c c c c|}
\hline
Model & $z$ & Method & $\delta_c$ & $\Delta_{\rm vir}$ & $k_\star$ [$ h\,\mathrm{Mpc}^{-1}$] & $\mathcal{E}$ \\
\hline
nDGPm & 0 & Our & 1.692 & 282.1 & 1.29 & 0.96 \\
nDGPm & 0 & \texttt{ReACT} & 1.691 & 282.6 & 1.29 & 0.96 \\
nDGPm & 1 & Our & 1.703 & 179.1 & 4.43 & 0.95 \\
nDGPm & 1 & \texttt{ReACT} & 1.703 & 178.4 & 4.41 & 0.95 \\
\hline
nDGPw & 0 & Our & 1.686 & 312.2 & 1.40 & 0.98 \\
nDGPw & 0 & \texttt{ReACT} & 1.685 & 312.6 & 1.39 & 0.98 \\
nDGPw & 1 & Our & 1.693 & 193.2 & 4.05 & 0.98 \\
nDGPw & 1 & \texttt{ReACT} & 1.693 & 192.7 & 4.05 & 0.98 \\
\hline
\end{tabular}
\label{tab:halo_parameters_full}
\end{table*}

\begin{figure}[h!]
    \centering
    \includegraphics[width=1\linewidth]{Final_Figures/compare_R_z0z1_nDGPm_nDGPw_kstar_large.png}
\caption{Comparison of the reaction for nDGPm ($r_c H_0 = 0.5$) and nDGPw ($r_c H_0 = 0.25$) computed with our code and \texttt{ReACT}. In our calculations, we use the SPT one-loop result from \texttt{ReACT} to determine $k_\star$, and also show the $k_\star \rightarrow \infty$ estimate. Both computations are compared with \texttt{ReACT} (dotted: $k_\star \rightarrow \infty$, solid: $k_\star$ from \texttt{ReACT}), where the relative difference is defined as $\Delta \mathcal{R} \equiv \mathcal{R}_{\mathrm{Our}}/\mathcal{R}_{\mathrm{ReACT}} - 1$. Left panel: $z=0$; right panel: $z=1$.}
    \label{fig:nDGPm_nDGPw}
\end{figure}

\begin{figure}[h!]
    \centering
    \includegraphics[width=0.5\linewidth]{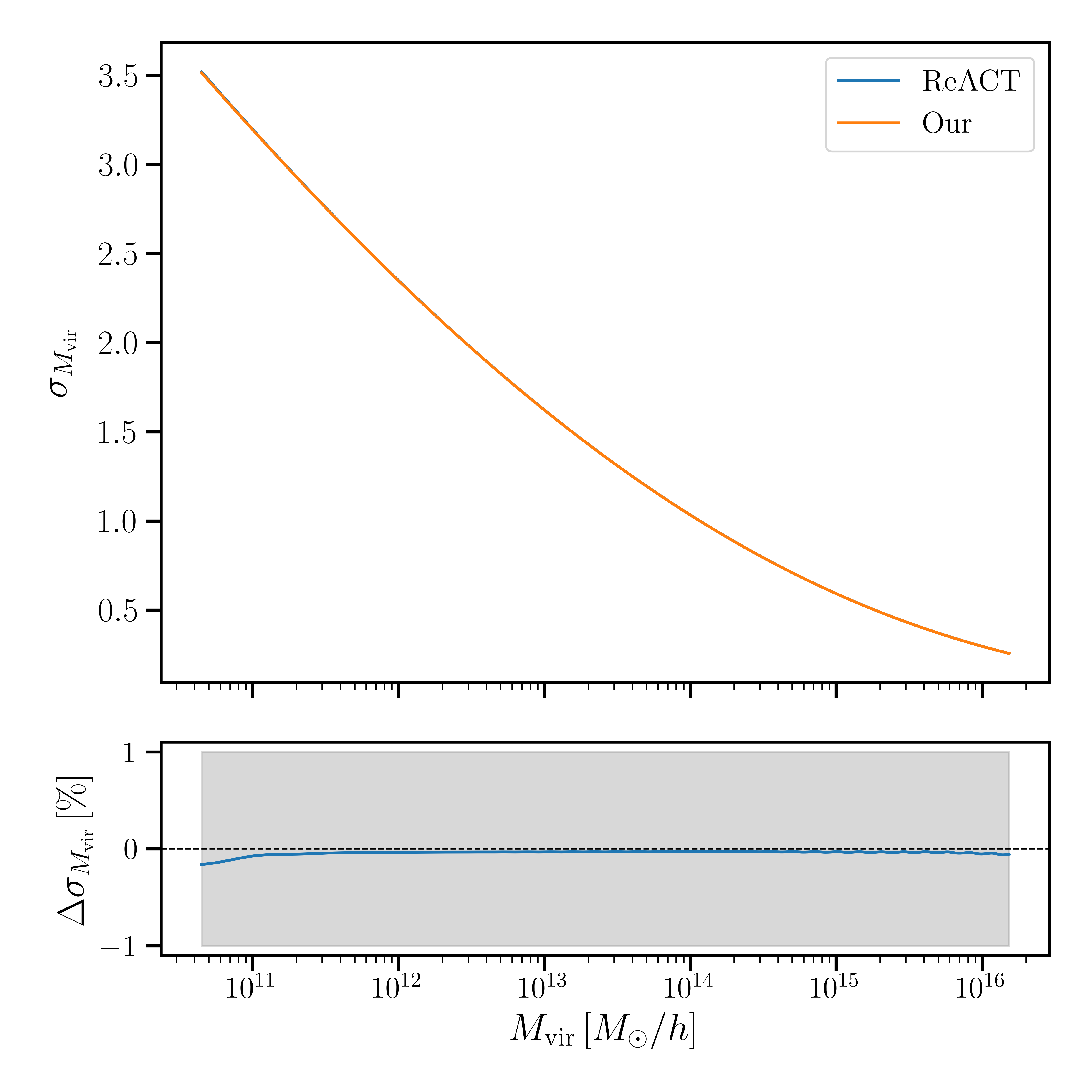}
    \caption{Variance $\sigma_{M_{\mathrm{vir}}}$, as a function of $M_{\mathrm{vir}}$ for the CG model at $z=0$, computed with our code and compared with \texttt{ReACT}. The lower panel shows the relative difference expressed in percent, highlighting deviations from \texttt{ReACT}.}
    \label{fig:sig_CG_z0}
\end{figure}

\begin{figure}[h!]
    \centering
    \includegraphics[width=0.9\linewidth]{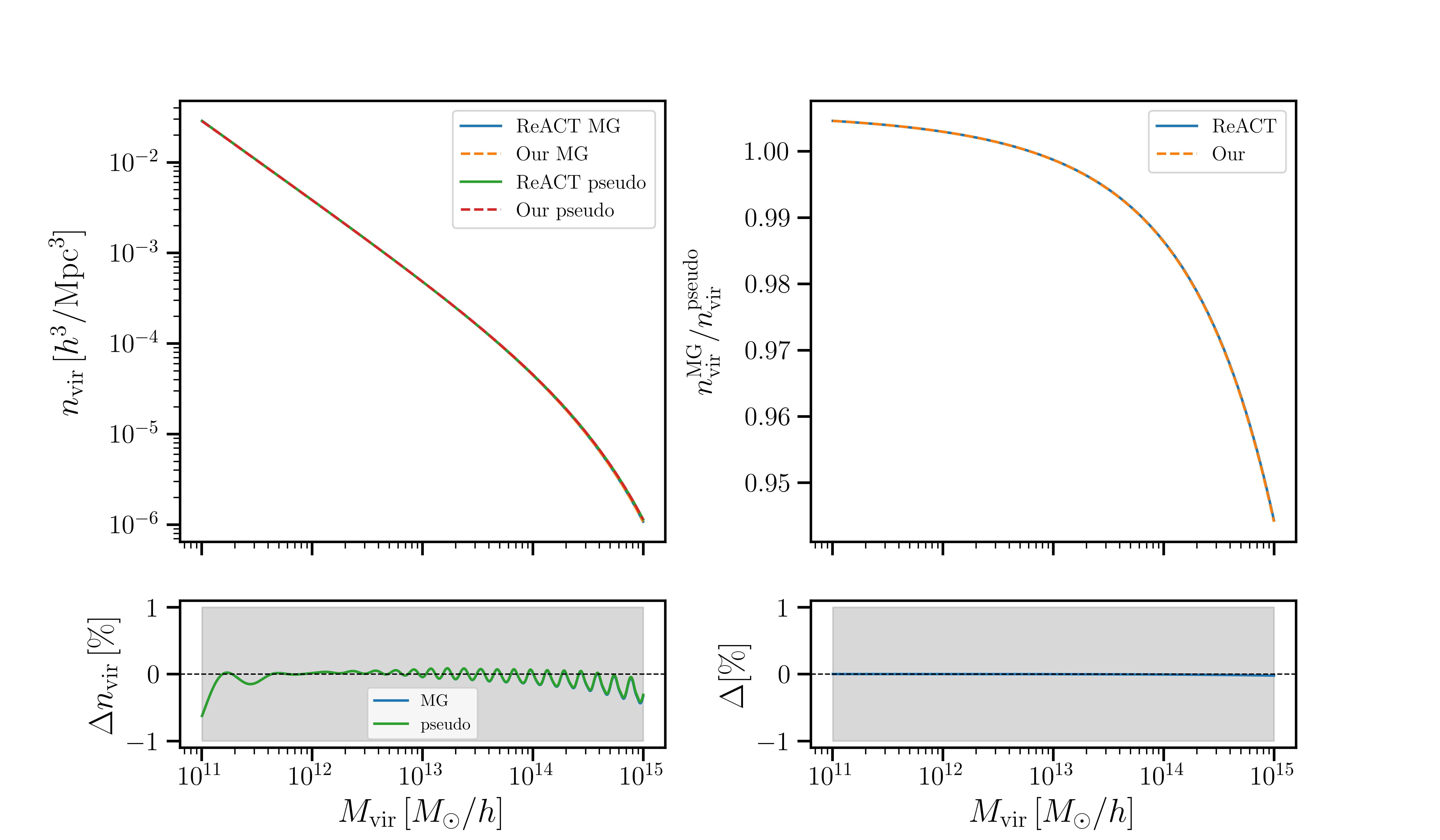}
    \caption{Left panel: comparison of the halo mass function $n_{\mathrm{vir}}$ of the CG model and its pseudo counterpart, computed with our code and \texttt{ReACT}, with the residuals shown below. Right panel: the ratio $n_{\mathrm{vir}}^{\mathrm{MG}} / n_{\mathrm{vir}}^{\mathrm{pseudo}}$, again computed with our code and \texttt{ReACT}, with residuals shown below. All quantities are evaluated at redshift $z=0$.}
    \label{fig:hmf_CG_z0}
\end{figure}

\begin{figure}[h!]
    \centering
    \includegraphics[width=0.85\linewidth]{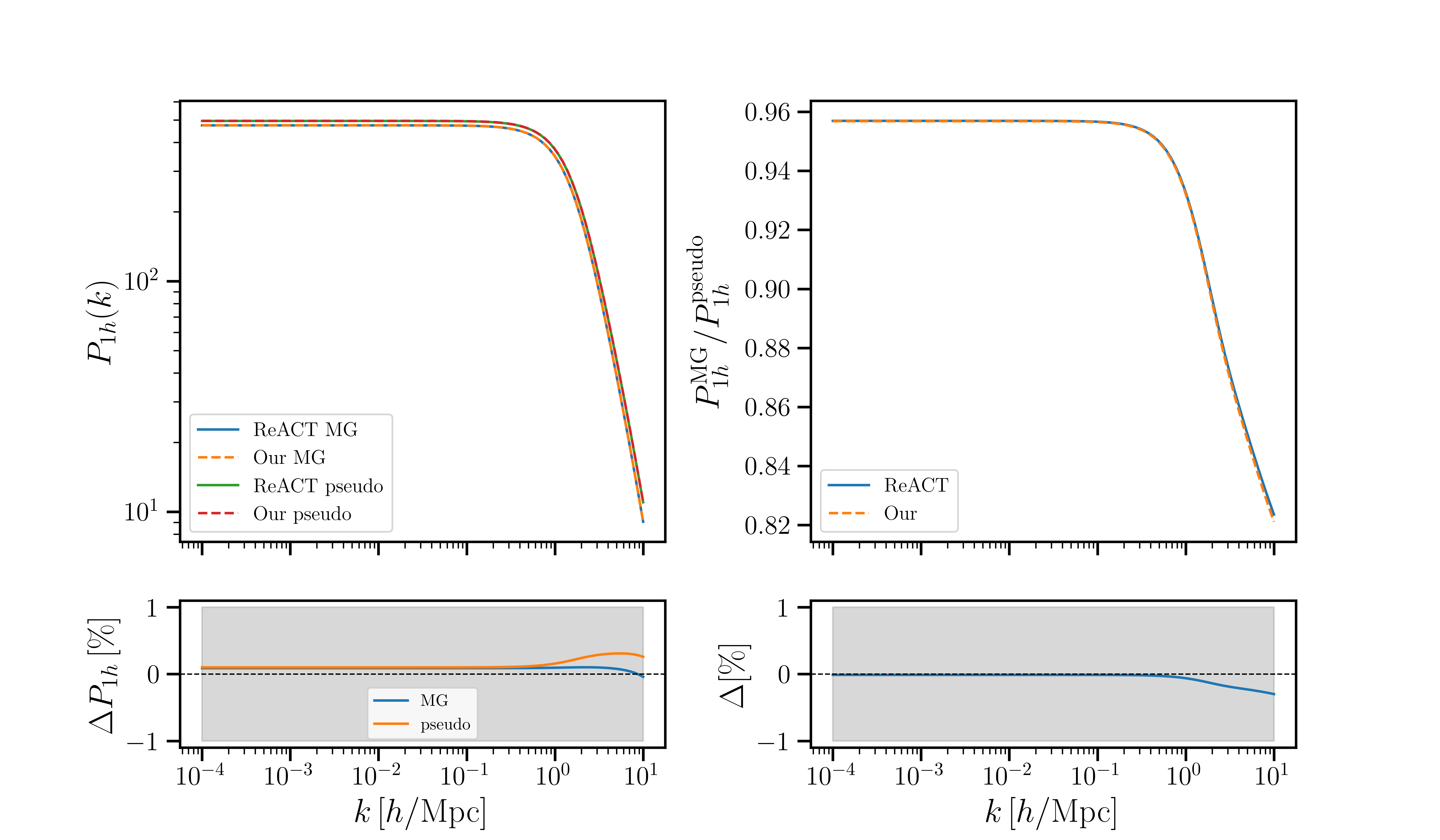}
    \caption{Left panel: comparison of the 1-halo term $P_{1h}$ for the CG model and its pseudo counterpart, computed with our code and \texttt{ReACT}, with the residuals shown below. Right panel: the ratio $P_{1h}^{\mathrm{MG}} / P_{1h}^{\mathrm{pseudo}}$, again computed with our code and \texttt{ReACT}, with residuals shown below. All quantities are evaluated at redshift $z=0$.}
    \label{fig:p1h_cg_z0}
\end{figure}

\begin{figure}[h!]
    \centering
    \includegraphics[width=0.85\linewidth]{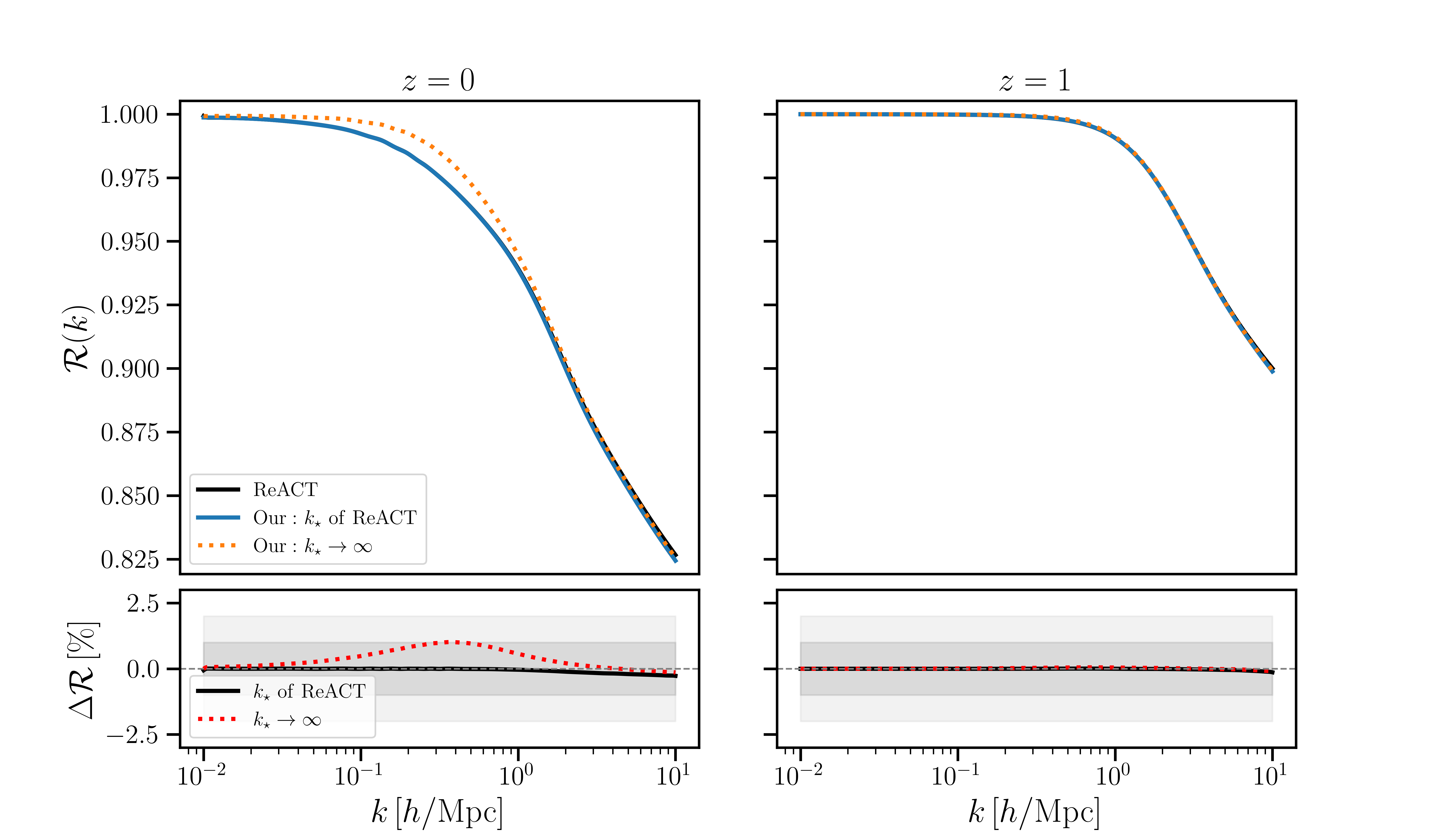}
    \caption{Comparison of the reaction for CG computed with our code and \texttt{ReACT}. In our calculations, we use the SPT one-loop result from \texttt{ReACT} to determine $k_\star$, and also show the $k_\star \rightarrow \infty$ estimate. Both computations are compared directly with \texttt{ReACT}. Left panel: $z=0$; right panel: $z=1$.}
    \label{fig:R_kstar_CG}
\end{figure}

We now move on to the Cubic Galileon (CG) model discussed in Ref.~\cite{Atayde:2024tnr}. This model is already implemented in \texttt{EFTCAMB} and naturally contained in our framework, corresponding to a specific choice of the $\alpha_i$ and Vainshtein screening. As such, with this comparison we effectively test all the ingredients of the halo-model.  For the halo-model ingredients we only display the comparison  at $z=0$, as the validation for $z=1$ gives similar results. For the reaction instead we show the results at both redshifts. In particular we investigate the following:
\begin{itemize}
    \item Spherical collapse and related parameters in Table~\ref{tab:halo_parameters}, where $k_\star$ is computed using the SPT result at $\bar{k}=0.06\, h\,\mathrm{Mpc}^{-1}$ from \texttt{ReACT},
    \item $\sigma_{M_{\mathrm{vir}}}$ of the pseudo linear power spectrum in Figure~\ref{fig:sig_CG_z0},
    \item $n_{\mathrm{vir}} \equiv dn/d\ln M_{\mathrm{vir}}$ for real and pseudo cosmologies (in the MPS formalism), and their ratio in Figure~\ref{fig:hmf_CG_z0},
    \item $P_{1h}$ for real and pseudo cosmologies, and their ratio in Figure~\ref{fig:p1h_cg_z0},
    \item $\mathcal{R}$ computed with $k_\star$ from \texttt{ReACT} and with $k_\star \rightarrow \infty$ in Figure~\ref{fig:R_kstar_CG}.
\end{itemize}
We find sub-percent agreement for all quantities.

\begin{table}[h!]
\caption{
Comparison of our and \texttt{ReACT} halo parameters at $z=0$ and $z=1$ for the CG model. 
Here, $\delta_c$ is the linear collapse threshold, $\Delta_{\rm vir}$ is the virial overdensity, 
$k_\star$ is the characteristic scale, and 
$\mathcal{E} = P_{1h}^{\mathrm{MG}} / P_{1h}^{\mathrm{pseudo}}(k \rightarrow 0)$. The \texttt{ReACT} values of $\delta_c$ have been converted to our MG convention, 
and $k_\star$ in our case is computed from the SPT result at $\bar{k} = 0.06 \, h\,\mathrm{Mpc}^{-1}$.}

\centering
\begin{tabular}{|c c c c c c|}
\hline
$z$ & Method & $\delta_c$ & $\Delta_{\rm vir}$ & $k_\star$ [$ h\,\mathrm{Mpc}^{-1}$] & $\mathcal{E}$ \\
\hline
0 & Our & 1.694 & 287.9 & 0.79 & 0.96 \\
0 & \texttt{ReACT} & 1.694 & 288.4 & 0.78 & 0.96 \\
1 & Our & 1.686 & 183.3 & 5.22 & 0.99 \\
1 & \texttt{ReACT} & 1.686 & 183.2 & 5.19 & 0.99 \\
\hline
\end{tabular}
\label{tab:halo_parameters}
\end{table}

Finally, we focus on EFTofDE models. By default, only $\alpha_i \propto a$ was implemented in \texttt{ReACT},  therefore we implemented the $\alpha_i \propto \Omega_{\mathrm{DE}}$ model ourselves. For the $\alpha_i \propto \Omega_{\rm DE}$ implementation in \texttt{ReACT} \footnote{For the public (unmodified) version of \texttt{ReACT}, see \url{https://github.com/nebblu/ACTio-ReACTio}.}, we modified the EFT time dependence in \texttt{reactions/src/BeyondLCDM.cpp}, replacing the default $\alpha_i(a)\propto a$ parameterization with $\alpha_i(a)\propto \Omega_{\rm DE}(a)$, assuming a $\Lambda$CDM background expansion.  We also consistently modified the computation of $M_\star$ and the corresponding time derivatives of the EFT functions. All other aspects of the implementation follow the default \texttt{ReACT} structure without modification. We do not plan to make this modification to \texttt{ReACT} publicly available, but we provide sufficient detail for the implementation to be reproduced. The EFTofDE screening prescriptions in \texttt{ReACT} differ from the Vainshtein-type nonlinear modification $\mu^{\mathrm{NL}}$ used in this work. To enable a consistent comparison, we validate our implementation against the "unscreened" case in \texttt{ReACT}, defined by $\mu^{\mathrm{NL}} = \mu^{\mathrm{L}}$, which we also implemented in our framework. \texttt{ReACT} assumes $k_\star = 1\, h\,\mathrm{Mpc}^{-1}$ and $\mathcal{E}=1$ for these "unscreened" EFTofDE models, which at the level of the reaction is equivalent to the limit $k_\star \rightarrow \infty$. We reiterate that the overlap test is performed in the unscreened case, where it tests the linear EFT evolution and reaction machinery, but does not test the Vainshtein-type nonlinear completion.

\begin{figure}[h!]
    \centering
    \includegraphics[width=0.9\linewidth]{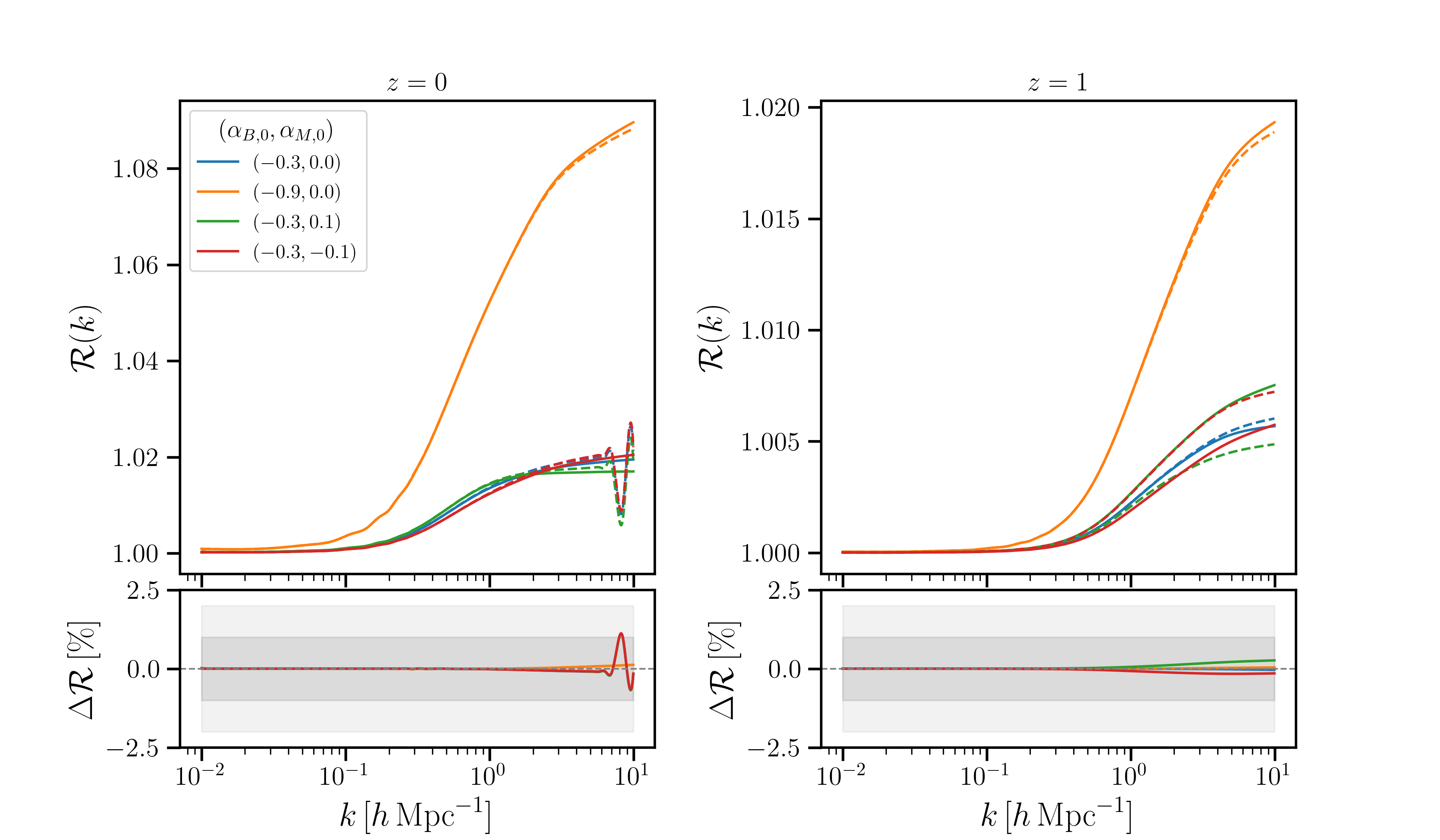}
    \caption{Reaction of EFTofDE models in the "unscreened" scenario ($\mu^{\rm L} = \mu^{\rm NL}$). Solid lines: our results; dashed lines: \texttt{ReACT}. Each curve corresponds to a model with $\alpha_i \propto \Omega_{\mathrm{DE}}$. Left panel: $z=0$; right panel: $z=1$. The $y$-axis ranges of the subplots are adjusted individually to ensure clear visibility at $z=1$.}
    \label{fig:alphaDE_un}
\end{figure}

In Fig.~\ref{fig:alphaDE_un} we present a comparison between the unscreened configuration of our implementation and \texttt{ReACT} at redshifts $z=0$ and $z=1$. Overall, we find agreement at the sub-percent level. Note that  in the deeply nonlinear regime, on scales well beyond the regime of validity of the halo model ($k \gtrsim 3 \, h\,\mathrm{Mpc}^{-1}$), \texttt{ReACT} exhibits small noisy features at $z=0$. These are numerical noise arising either from our specific implementation of $\alpha_i \propto \Omega_{\mathrm{DE}}$ within \texttt{ReACT} or from internal numerical effects of \texttt{ReACT}. In any case, these features do not affect the conclusions of this comparison.

As a side note, one can see that these unscreened $\alpha_i\propto\Omega_{\rm DE}$  models have a reaction which is larger than unity. When instead accounting for the (Vainshtein) screening in our main analysis we found a reaction smaller than unity (see e.g., Fig.~\ref{fig:Ri_vary_kstar}). This shows that the screening in fact leads to a suppression of large-scale structure with respect to \LCDM{}.

\begin{table*}[htbp]
\caption{
MG spherical collapse parameters for the unscreened EFTofDE models.}
\centering
\setlength{\tabcolsep}{5pt}
\begin{tabular}{|l c c c c|}
\hline
$(\alpha_{B,0},\alpha_{M,0})$ 
& $\delta_c(z=0)$ 
& $\delta_c(z=1)$ 
& $\Delta_{\rm vir}(z=0)$ 
& $\Delta_{\rm vir}(z=1)$ \\
\hline

$(-0.3,0)$ \quad Our
& 1.669 & 1.683 & 340.8 & 201.2 \\
$(-0.3,0)$ \quad \texttt{ReACT}
& 1.669 & 1.683 & 340.6 & 201.1 \\

\hline

$(-0.9,0)$ \quad Our
& 1.653 & 1.680 & 365.3 & 204.2 \\
$(-0.9,0)$ \quad \texttt{ReACT}
& 1.653 & 1.680 & 364.9 & 203.9 \\

\hline

$(-0.3,0.1)$ \quad Our
& 1.669 & 1.683 & 339.4 & 201.4 \\
$(-0.3,0.1)$ \quad \texttt{ReACT}
& 1.671 & 1.683 & 336.0 & 200.8 \\

\hline

$(-0.3,-0.1)$ \quad Our
& 1.670 & 1.683 & 341.5 & 201.0 \\
$(-0.3,-0.1)$ \quad \texttt{ReACT}
& 1.667 & 1.683 & 344.3 & 201.5 \\

\hline
\end{tabular}
\end{table*}

\clearpage
\bibliographystyle{JHEP}
\bibliography{references}

\end{document}